\documentclass[preprint,12pt]{elsarticle}

 \usepackage{amssymb, latexsym}
 \usepackage{amsmath} 
 \usepackage{amsthm}
 \usepackage{graphics}
 \usepackage{epstopdf}
 \usepackage{epsf} 
 \usepackage{mwe}    % loads »blindtext« and »graphicx«
 \usepackage{subfig}
 \usepackage{bm}
 \usepackage[mathscr]{eucal}
 \usepackage{enumerate}

\usepackage{amssymb}
\journal{Journal of Sound and Vibration}

\begin{document}

\begin{frontmatter}



\title{An integral formulation for wave propagation on weakly non-uniform potential flows}


\author{Simone Mancini\footnote{e-mail: s.mancini@soton.ac.uk}, R. Jeremy Astley, Samuel Sinayoko and Gw\'ena\"el Gabard}

\address{Institute of Sound and Vibration Research,	University of Southampton, Southampton, United Kingdom - SO17 1BJ}

\author{Michel Tournour}

\address{Siemens Industry Software NV, Interleuvenlaan 68, 3001, Leuven, Belgium}

\begin{abstract}
	
 An integral formulation for acoustic radiation in moving flows is presented. It is based on a potential formulation for acoustic radiation on weakly non-uniform subsonic mean flows. This work is motivated by the absence of suitable kernels for wave propagation on non-uniform flow. The integral solution is formulated using a Green's function obtained by combining the Taylor and Lorentz transformations. Although most conventional approaches based on either transform solve the Helmholtz problem in a transformed domain, the current Green's function and associated integral equation are derived in the physical space. A dimensional error analysis is developed to identify the limitations of the current formulation. Numerical applications are performed to assess the accuracy of the integral solution. It is tested as a means of extrapolating a numerical solution available on the outer boundary of a domain to the far field, and as a means of solving scattering problems by rigid surfaces in non-uniform flows. The results show that the error associated with the physical model deteriorates with increasing frequency and mean flow Mach number. However, the error is generated only in the domain where mean flow non-uniformities are significant and is constant in regions where the flow is uniform.

\end{abstract}

%\begin{keyword}
%%% keywords here, in the form: keyword \sep keyword
%wave propagation \sep boundary integral formulation \sep non-uniform mean flow \sep Taylor-Lorentz transform \sep wave extrapolation \sep boundary element method
%%% PACS codes here, in the form: \PACS code \sep code
%
%%% MSC codes here, in the form: \MSC code \sep code
%%% or \MSC[2008] code \sep code (2000 is the default)
%
%\end{keyword}

\end{frontmatter}


\section{Introduction}\label{sec:introduction}
%--------------------------------------------------------------------------------------------
Predicting noise radiation from complex sources in moving flows is relevant to the automotive, energy and aeronautical industries. Noise radiation from turbofan nacelles and from other aircraft sources is a problem of particular interest in the aviation sector. Numerical simulation of noise radiation and scattering can significantly reduce costs for design and certification. However, an efficient numerical method for high frequency noise propagation on non-uniform moving flows has not yet been demonstrated. Solving high frequency short wavelength problems on moving flows remains computationally expensive. In the aeronautical industry, noise propagation on non-uniform flows is typically predicted using finite element methods (FEM)~\cite{Astley1981}, discontinuous Galerkin methods (DGM)~\cite{Gabard2007} and high order finite difference schemes~\cite{Tam1993}.

Although volume based methods, such as FEM, DGM and finite difference schemes are able to solve wave propagation on a non-uniform flow, predicting noise radiation in unbounded domain requires the computational domain to be truncated. The truncation of the domain allows acoustic waves to be damped in a non-physical absorbing zone~\cite{Tam1993,Eversman1999,Bermudez2007} and satisfy the radiation condition at the outer boundary of the domain. Moreover, these methods suffer of dispersion error and pollution effects~\cite{Babuska1997}. These features are relevant limitations in case of noise radiation for large-scale short-wavelength problems.

On the other hand, numerical methods based on boundary integral formulations, such as the boundary element method (BEM)~\cite{Wu2000}, inherently satisfy the radiation condition in the kernel and allow wave propagation in unbounded domains to be solved more effectively than in the case of volume based methods. Moreover, the fast multiple BEM (FMBEM) is an efficient algorithm to solve wave radiation and scattering for large-scale short-wavelength problems~\cite{Delnevo2005}. However, BEM can only solve wave propagation exactly on uniform mean flows. Extending this method to non-uniform flow regions would be beneficial to a number of applications, such as forward fan noise acoustic installation effects. A surface integral formulation including non-uniform flow effects would also extend the applicability of wave extrapolation methods. These approaches use an integral formulation defined on a closed surface on which the acoustic field is sampled from an `inner' domain to radiate the solution to the far field. At the moment these methods are limited to uniform flow~\cite{FfowcsWilliams1969,Farassat1988}.

Current boundary element modelling practices use the Lorentz transformation~\cite{Chapman2000, Gregory2014} to solve wave propagation on a uniform mean flow. This variable transformation allows the uniform flow Helmholtz equation to be reduced to the standard Helmholtz problem without approximations. By means of a Lorentz transformation, BEM solvers for the standard Helmholtz equation can therefore be used for wave propagation on uniform flows. However, due to the variable transformation, the physical space is deformed in the direction of the mean flow. The deformation of the domain complicates the formulation of the boundary conditions and the implementation of the transmission conditions for coupled formulations~\cite{Balin2014,Casenave2014}. Alternatively, this drawback can be overcome by using an integral formulation in the physical space as proposed by Wu and Lee~\cite{Wu1994} in the frequency domain and by Hu \cite{Hu2013} in the time domain.

For BEM, only approximate formulations are available for representing non-uniform mean flow effects. Astley and Bain~\cite{Astley1986} provided an approximate formulation for wave propagation on low Mach number mean flows based on Taylor's transformation~\cite{Taylor1979, Agarwal2007}. In the same work, Astley and Bain~\cite{Astley1986} reported an error analysis for Taylor's wave equation showing that the accuracy of the physical model depends only upon the mean flow Mach number and the characteristic length scales of the acoustic waves and the mean flow. On the other hand, Tinetti and Dunn~\cite{Tinetti2005} provided a generalized local Lorentz transformation to represent the effect of non-uniform mean flows on wave propagation. However, the method has been restricted to mean flow fields with small gradients. Another approach is to move the terms including non-uniform flow effects to the right hand side of the equation and treat them as sources in the domain. The dual-reciprocity method (DRM)~\cite{Lee1994} is then used to convert the domain integrals into boundary integrals. The absence of a robust method to define interpolating source functions for the DRM restricts the applicability of this approach. Thereby, modeling non-uniform flow effects for BEM is still an open problem.  

In this article, we present, in the physical space, an integral formulation with non-uniform flow based on a combination of the physical models associated with the Taylor and Lorentz transformations. The proposed physical model is an approximate formulation of the full linearized potential wave equation for isentropic compressible flows. The integral formulation derived applies to sound radiation on a \emph{weakly non-uniform} potential mean flow. Consider the mean flow as a sum of a uniform and a non-uniform component which vanishes at infinity. The term \emph{weakly non-uniform} indicates that the non-uniform portion of the mean flow is small compared to the uniform part. A free field Green's function is also determined for a subsonic weakly non-uniform flow as a kernel for the integral equation. Moreover, an error analysis is presented to extend and to revisit the estimate provided by Astley and Bain~\cite{Astley1986}. This analysis shows the dependency of the error related to the physical model on the mean flow Mach number and on the frequency. The proposed formulation will be shown to improve the accuracy of the model compared to existing formulations based on either Taylor or Lorentz transforms being applied separately. 

The paper is structured as follows. Section~\ref{sec:physical_model} presents the physical models. In Section~\ref{sec:integral_formuation_small_perturbation_convected_wave_equation} integral formulations are derived for wave propagation on weakly non-uniform mean flows and the Taylor formulation in the physical space. The Green's functions associated with the integral formulations are derived in Section~\ref{sec:Green_function_adjoint_operator}. Boundary element formulations consistent with the proposed integral solutions are presented in Section~\ref{sec:variational_formulation_small_perturbation_convected_wave_equation_with_source_terms} for an arbitrary source distribution. In Section~\ref{sec:Error_Estimate_small_perturbation_convected_wave_equation}, a dimensional error analysis is developed to describe the limitation of the proposed solutions. Finally, in Section~\ref{sec:numerical_results_BIF} some numerical results are presented to benchmark the integral formulations. 

\section{Physical model} \label{sec:physical_model}
%--------------------------------------------------------------------------------------------
\begin{figure}[t]
	\centering
	\includegraphics[width=0.9\textwidth]{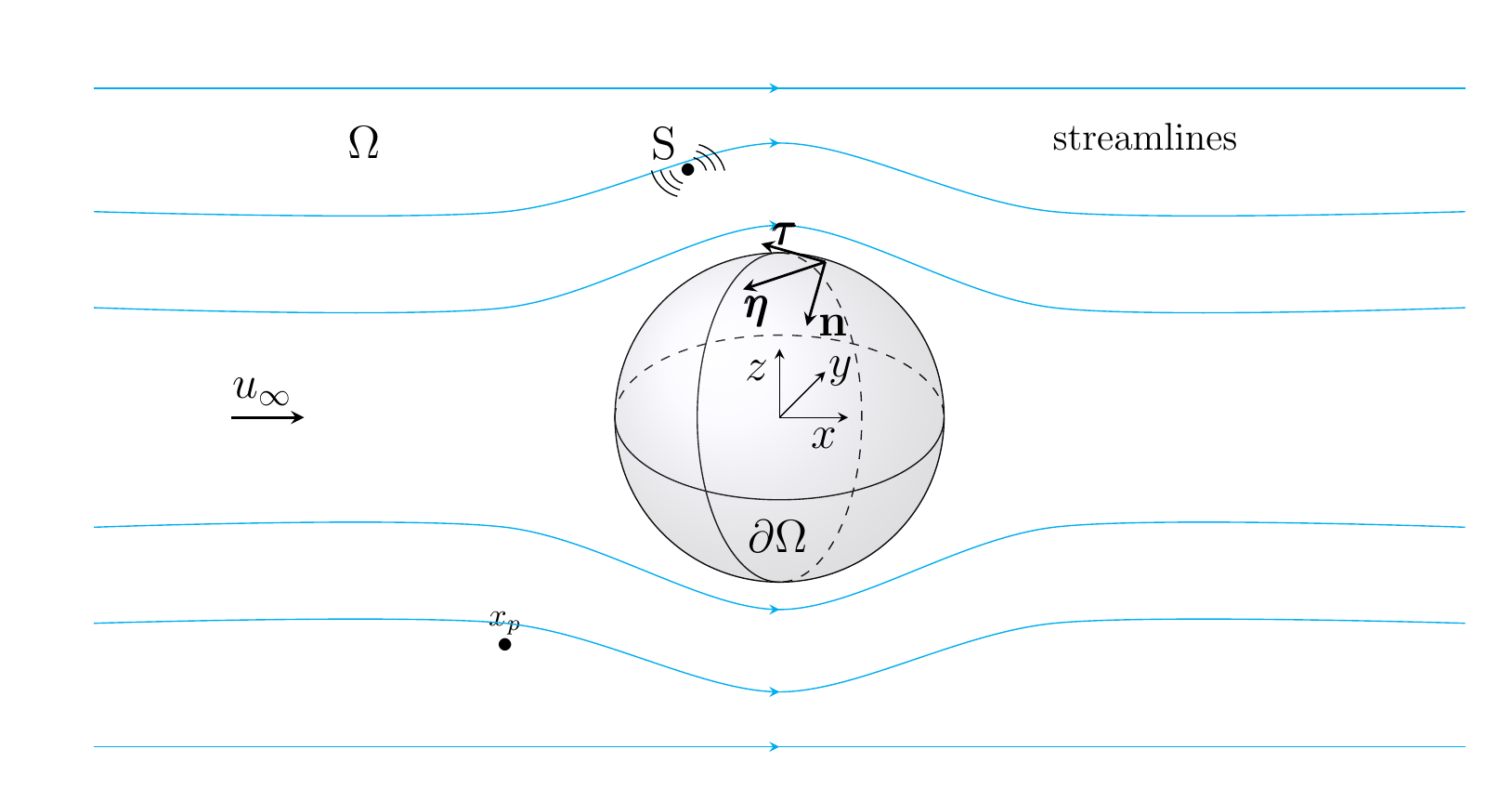} 
	\caption{\footnotesize Schematic diagram of the reference problem showing mean flow streamlines in the solution domain $\Omega$.}
	\label{fig:geometry_Domain}
\end{figure}

A numerical solution to external noise radiation and scattering of a sound field from a source $S$ in a domain $\Omega$ by a body $\partial \Omega$ in a non-uniform potential subsonic mean flow is sought (see Fig.~\ref{fig:geometry_Domain}).
Consider an inviscid, adiabatic and irrotational flow and assume that acoustic perturbations are of small amplitude compared with the steady mean flow. Under these hypotheses, wave propagation on a non-uniform mean flow can be described by means of a potential formulation~\cite{Astley1985} as,
\begin{equation}
\frac{D_0}{D t}\left(\frac{\rho_0}{c_0^2}\frac{D_0 \hat \phi}{D t}\right)-{\pmb{\nabla}}\cdot\left(\rho_0{\pmb{\nabla}}\hat \phi\right)=0
\label{eq:full_potential_acoustics}
\end{equation}
where $\hat \phi$ is the acoustic velocity potential and $D_0/Dt= \partial/\partial t + {\bf{u}}_0\cdot {\pmb{\nabla}}$ denotes the material derivative over the mean flow; $\rho_0$ is the mean flow density, $c_0$ the speed of sound and ${\bf{u}}_0$ the mean flow velocity. Consider Bernoulli's equation
\begin{equation}
c_0^2=c^2_\infty-\frac{\gamma-1}{2}(\|{\bf{u}}_0\|^2-\|{\bf{u}}_\infty\|^2)
\label{eq:speed_of_sound_insentropic}
\end{equation}
where $c_\infty$ and ${\bf{u}}_\infty$ are the speed of sound and the mean flow velocity vector in the far field. By means of Eq.~(\ref{eq:speed_of_sound_insentropic}) and introducing the state equation,
\begin{equation}
\frac{d \rho_0}{\rho_0}=\frac{\gamma-1}{2}\frac{d c_0^2}{c_0^2},
\label{eq:drho_insentropic}
\end{equation}
Eq.~(\ref{eq:full_potential_acoustics}) reduces to:
\begin{equation}
\begin{split}
\frac{\partial^2 \hat \phi}{\partial t^2}+& 2{\bf{u}}_0\cdot{\pmb{\nabla}}\frac{\partial\hat  \phi}{\partial t}-c_\infty^2\nabla^2\hat \phi+{\bf{u}}_0\cdot{\pmb{\nabla}}({\bf{u}}_0\cdot{\pmb{\nabla}}\hat \phi)\\
&+\frac{1}{2}{\pmb{\nabla}}\hat \phi\cdot{\pmb{\nabla}}({\bf{u}}_0\cdot{\bf{u}}_0)+(\gamma-1)\frac{D_{0} \hat \phi}{D t}{\pmb{\nabla}}\cdot{\bf{u}}_0\\
&+\frac{\gamma -1}{2}(\|{\bf{u}}_0\|^2-\|{\bf{u}}_\infty\|^2)\nabla^2\hat \phi= 0 .\\
\end{split}
\label{eq:full_potential_acoustics_v2}
\end{equation}

Equation~(\ref{eq:full_potential_acoustics}) is rewritten as Eq.~(\ref{eq:full_potential_acoustics_v2}) to allow a dimensional analysis to be performed.
Mayoral and Papamoschou~\cite{Mayoral2013}, following Astley and Bain~\cite{Astley1986}, have provided a description of the dependency of the terms in Eq.~(\ref{eq:full_potential_acoustics_v2}) on the mean flow Mach number $M_\infty=u_\infty/c_\infty$, the acoustic characteristic length scale $L_A$ and the characteristic length scale of the mean flow $L_M$. On this basis, Eq.~(\ref{eq:full_potential_acoustics_v2}) is simplified retaining only terms of order $[\phi]/L^2_A$, $M_\infty[\phi]/L^2_A$, $M_\infty^2[\phi]/L^2_A$ and $M_\infty^2[\phi]/L_AL_M$ to give
\begin{equation}
\begin{split}
\frac{\partial^2 \hat \phi}{\partial t^2}+& 2{\bf{u}}_0\cdot{\pmb{\nabla}}\frac{\partial\hat  \phi}{\partial t}-c_\infty^2\nabla^2\hat \phi+{\bf{u}}_0\cdot{\pmb{\nabla}}({\bf{u}}_0\cdot{\pmb{\nabla}}\hat \phi)\\
&+\frac{1}{2}{\pmb{\nabla}}\hat \phi\cdot{\pmb{\nabla}}({\bf{u}}_0\cdot{\bf{u}}_0)+\frac{\gamma -1}{2}(\|{\bf{u}}_0\|^2-\|{\bf{u}}_\infty\|^2)\nabla^2\hat \phi= 0 .\\
\label{eq:full_potential_acoustics_approx1}
\end{split}
\end{equation}
Assume that $L_A\leq L_M$ and $M_\infty\ll 1$. Then consider the mean flow formed by a uniform component ${\bf{u}}_\infty$ and a non-uniform portion ${{\bf{u}}'_0}$. The main idea is to retain only first order effects in ${{\bf{u}}'_0}$ over wave convection due to ${\bf{u}}_\infty$. 
If ${\bf{u}}_0 = {\bf{u}}_\infty+{{\bf{u}}'_0}$, where $\|{{\bf{u}}'_0}\|\ll\|{\bf{u}}_\infty\|$ and if it is assumed that the uniform mean flow velocity ${\bf{u}}_\infty$ is aligned with the positive $x$-axis, Eq.~(\ref{eq:full_potential_acoustics_approx1}) reduces to:
\begin{equation}
\frac{\partial^2 \hat \phi}{\partial t^2}+2{\bf{u}}_0\cdot{\pmb{\nabla}}\frac{\partial \hat \phi}{\partial t}-c_\infty^2\nabla^2\hat \phi+u_\infty^2\frac{\partial^2\hat  \phi}{\partial x^2} =0 .
\label{eq:convected_small_distur_wave_eq}
\end{equation}
Hereafter, Eq.~(\ref{eq:convected_small_distur_wave_eq}) is referred to as the wave equation for \emph{weakly non-uniform} potential flow.

In the case of a uniform flow, where ${\bf{u}}_0\equiv{\bf{u}}_\infty$ at all points in the flow, Eq.~(\ref{eq:convected_small_distur_wave_eq}) reduces to the uniform flow wave equation without approximation,
\begin{equation}
\Bigg(\frac{\partial}{\partial t}+u_\infty\frac{\partial}{\partial x}\Bigg)^2 \hat \phi - c^2_\infty \nabla^2 \hat \phi =0 .
\label{convected_wave_equation}
\end{equation}
If the mean flow is uniform, solutions of Eq.~(\ref{convected_wave_equation}) are also solution of Eq.~(\ref{eq:full_potential_acoustics}) without approximation.
On the other hand, if terms of the order $M^2_\infty$ are neglected in Eq.~(\ref{eq:convected_small_distur_wave_eq}), it reduces to
\begin{equation}
\frac{\partial^2 \hat \phi}{\partial t^2}+2{\bf{u}}_0\cdot{\pmb{\nabla}}\frac{\partial \hat \phi}{\partial t}-c_\infty^2\nabla^2\hat \phi=0 .
\label{eq:low_Mach_number_wave_eq}
\end{equation}
In all that follows, Equation~(\ref{eq:low_Mach_number_wave_eq}) is referred to as the \emph{Taylor wave equation}, because it is consistent with the solution based on the Taylor transformation~\cite{Astley1986,Taylor1979}. It is an approximation of order $M_\infty$ of Eq.~(\ref{eq:full_potential_acoustics}) accounting for a non-uniform mean flow.

If a time harmonic problem in which all perturbed quantities varies as $\mathrm{e}^{\mathrm{i}\omega t}$ is considered, the acoustic velocity potential in the frequency domain can be defined as $\hat \phi = \phi e^{\mathrm{i}\omega t}$ where $\phi$ is a complex amplitude. In the Fourier domain Eqs.~(\ref{eq:convected_small_distur_wave_eq}), (\ref{convected_wave_equation}) and (\ref{eq:low_Mach_number_wave_eq}) are respectively referred to as the \emph{weakly non-uniform potential flow Helmholtz equation}, the \emph{uniform flow Helmholtz equation} and the \emph{Taylor-Helmholtz equation}.

\section {Boundary integral formulation}\label{sec:integral_formuation_small_perturbation_convected_wave_equation}

\subsection {Weakly non-uniform potential flow Helmholtz equation}\label{sec:weakly_non_unif_bondary_integral_solution}

By assuming a harmonic time dependence for $\hat \phi$, Eq.~(\ref{eq:convected_small_distur_wave_eq}) can be rewritten
\begin{equation}
k^2\phi-2\mathrm{i}k{\bf{M}}_0\cdot {\pmb{\nabla}} \phi + \nabla^2\phi-M_\infty^2\frac{\partial^2 \phi}{\partial x^2} =0 ,
\label{eq:convected_small_distur_Helmholtz_eq}
\end{equation}
where ${\bf{M}}_0$ is the mean flow Mach number vector, $k=\omega/c_\infty$, and $M_\infty$ is the uniform flow Mach number, provided that ${\bf{M}}_\infty=(M_\infty,0,0)$. First, the reverse flow operator associated with Eq.~(\ref{eq:convected_small_distur_Helmholtz_eq}) is defined considering a mean flow in the opposite direction of the actual flow field to give
\begin{equation}
k^2\phi+2ik{\bf{M}}_0\cdot{\pmb{\nabla}} \phi + \nabla^2\phi-M_\infty^2\frac{\partial^2 \phi}{\partial x^2} =0.
\label{eq:reverse_flow_weakly_non_unifor_flow_Helmholtz_no_source}
\end{equation}

Following the general approach of Wu and Lee~\cite{Wu1994}, a Green's function $G$ for the fundamental reverse flow operator Eq.~(\ref{eq:reverse_flow_weakly_non_unifor_flow_Helmholtz_no_source}) satisfies
\begin{equation}
k^2G+2\mathrm{i}k{\bf{M}}_0\cdot {\pmb{\nabla }} G + \nabla^2G-M_\infty^2\frac{\partial^2 G}{\partial x^2} = -\delta(\textbf{x}_p-\textbf{x}),
\label{eq:convected_small_distur_Helmholtz_adjoint_operatror_eq}
\end{equation}
where $\textbf{x}_p$ is an arbitrary field point, $\textbf{x}$ denotes the point source location and $\delta$ is the Dirac delta function.
The Green's function $G$ represents the effect of a point source in a non-uniform flow whose direction is opposite to the actual base flow. $G$ therefore differs from the Green's function of the direct fundamental operator and is derived in Section~\ref{sec:Green_function_adjoint_operator}.

To obtain an integral formulation for the weakly non-uniform potential flow Helmholtz equation, Eq.~(\ref{eq:convected_small_distur_Helmholtz_eq}) is multiplied by $G$ and Eq.~(\ref{eq:convected_small_distur_Helmholtz_adjoint_operatror_eq}) by $\phi$. Subtracting these two equations and integrating over the domain $\Omega$ yields
\begin{equation}
\begin{split}
&\int_{\Omega} \phi\left( k^2G+2\mathrm{i}k{\bf{M}}_0\cdot{\pmb{\nabla}} G + \nabla^2G-M_\infty^2\frac{\partial^2 G}{\partial x^2}\right)dV\\
&-\int_{\Omega} G\left(k^2\phi-2\mathrm{i}k{\bf{M}}_0\cdot{\pmb{\nabla }}\phi + \nabla^2\phi-M_\infty^2\frac{\partial^2 \phi}{\partial x^2}\right)dV\\
&= -\int_{\Omega}\phi \delta(\textbf{x}_p-\textbf{x})dV.
\end{split}
\label{eq:weighted_Residual_formulation_Ga}
\end{equation}
The divergence theorem is applied to the l.h.s. of Eq.~(\ref{eq:weighted_Residual_formulation_Ga}) in order to obtain an integral on the boundary surface of the domain $\partial \Omega$. Consistent with the assumptions already made in the derivation of Eq.~({\ref{eq:convected_small_distur_wave_eq}), some terms of higher order in $M_\infty$ will be neglected. Consider the linear term in ${\bf{M}}_0$ in Eq.~(\ref{eq:weighted_Residual_formulation_Ga}),
	\begin{equation}
	\begin{split}
	&\int_{\Omega} (2\mathrm{i}k\phi {\bf{M}}_0\cdot{\pmb{\nabla }}G + 2ikG{\bf{M}}_0\cdot{\pmb{\nabla}} \phi )dV\\
	&= \int_{\Omega} 2\mathrm{i}k [{\pmb{\nabla}} \cdot({\bf{M}}_0 G \phi) - G \phi{\pmb{\nabla}}\cdot{\bf{M}}_0 ]dV.\\
	\end{split}
	\label{eq:integral_eq_nonuniflow_linear_term_Der}
	\end{equation}
	Since~\cite{Astley1986}
	\begin{equation}
	\nabla \cdot {\bf{M}}_0 = \frac{\frac{1}{2}{\bf{M}}_0 \cdot {\pmb{\nabla}} ({\bf{M}}_0\cdot{\bf{M}}_0)}{1-\frac{\gamma -1}{2}(M_0^2-M_\infty^2)},
	\label{eq:div_M0_equation}
	\end{equation}
	where $M_0 = \|{\bf{M}}_0\|$,
	is of the same order as $M_\infty^3$, and since high order terms in $M_\infty$ have already been neglected in Eq.~(\ref{eq:convected_small_distur_wave_eq}), the second term on the r.h.s of Eq.~(\ref{eq:integral_eq_nonuniflow_linear_term_Der}) can be dropped, giving
	\begin{equation}
	\int_{\Omega} 2\mathrm{i}k [{\pmb{\nabla}} \cdot({\bf{M}}_0 G \phi) - G \phi{\pmb{\nabla}}\cdot{\bf{M}}_0 ]dV \simeq \int_{\partial \Omega} 2\mathrm{i}k{\bf{M}}_0\cdot {\bf{n}} G \phi dS,
	\end{equation}
	where ${\bf{n}}$ is the normal unit vector to $\partial \Omega$ (see Fig.~\ref{fig:geometry_Domain}). Hence, Eq.~(\ref{eq:weighted_Residual_formulation_Ga}) can be rewritten as
	\begin{equation}
	\begin{split}
	\phi({\bf{x}}_p) &= \int_{\partial \Omega} \left[G \frac{\partial \phi}{\partial n}- \phi\frac{\partial G}{\partial n}\right]dS \\
	&- \int_{\partial \Omega} \left[ 2\mathrm{i}k{\bf{M}}_0\cdot {\bf{n}} G \phi + M^2_\infty\left(G \frac{\partial \phi}{\partial x}-\phi \frac{\partial G}{\partial x}\right)n_x\right] dS,
	\end{split}
	\label{eq:Non_uniform_boundary_integral_domain}
	\end{equation}
	where $n_x$ is the $x-$component of the normal vector to the boundary surface. As shown by Wu and Lee~\cite{Wu1994} the contribution of the boundary integral at infinity is zero provided that $\phi({\bf{x}})$ and $G_0({\bf{x}}_p,{\bf{x}})$ satisfy the Sommerfeld radiation condition with flow.

	Equation~(\ref{eq:Non_uniform_boundary_integral_domain}) applies to any point ${\bf{x}}_p$ internal to the domain $\Omega$. On the other hand, for any point on the boundary surface $\partial \Omega$, a limit approach~\cite{Wu2000} should be used to overcome the singularity of the integral formulation, which is integrable in the sense of Cauchy's principal value. The following equation extends the uniform flow solution of Wu and Lee\cite{Wu1994} and is derived in~\ref{sec:Integral_form_along_bound_surface},
	\begin{equation}
	\begin{split}
	\hat C({\bf{x}}_p)\phi({\bf{x}}_p) &= \int_{\partial \Omega} \left[G \frac{\partial \phi}{\partial n}- \phi\frac{\partial G}{\partial n}\right] dS \\
	&-\int_{\partial \Omega} \left[ 2\mathrm{i}k{\bf{M}}_0\cdot {\bf{n}} G \phi + M^2_\infty\left(G \frac{\partial \phi}{\partial x}-\phi \frac{\partial G}{\partial x}\right)n_x\right] dS ,
	\end{split}
	\label{eq:Non_uniform_boundary_integral_surface_v2}
	\end{equation}
	where
	\begin{equation}
	\hat C({\bf{x}}_p)=
	\left\{
	\begin{array}{lr}
	1 & {\bf{x}}_p\in\Omega \\
	1 -\int_{\partial \Omega}\left(\frac{\partial G_0}{\partial n}-M_\infty^2\frac{\partial G_0}{\partial x}n_x \right) dS & \quad \quad {\bf{x}}_p\in\partial\Omega\\
	\end{array}\right.
	\end{equation} 
	and $G_0({\bf{x}}_p,{\bf{x}})$ is the Green's function associated with the static operator defined in~\ref{sec:Integral_form_along_bound_surface}.

	In summary, Eq.~(\ref{eq:Non_uniform_boundary_integral_surface_v2}) provides a boundary integral formulation for wave propagation on weakly non-uniform potential mean flows. It models first order non-uniform mean flow effects on wave propagation. The accuracy of the formulation depends on the deviation of the non-uniform flow ${\bf{M}}_0'$ on the uniform component ${\bf{M}}_\infty$. Equation~(\ref{eq:Non_uniform_boundary_integral_surface_v2}) is exact for wave propagation over a uniform mean flow and equivalent to the formulation presented by Wu and Lee~\cite{Wu1994} for a uniform flow. It reduces to the standard Helmholtz integral equation as $M_0 \rightarrow 0$.

\subsection{Taylor-Helmholtz equation}\label{sec:integral_formuation_low_Mach_number_convected_wave_equation}

For the Taylor-Helmholtz equation,
\begin{equation}
k^2\phi-2\mathrm{i}k{\bf{M}}_0\cdot {\pmb{\nabla}}  \phi + \nabla^2\phi =0 ,
\label{eq:low_mach_number_Helmholtz_eq}
\end{equation}
an integral solution in the frequency domain is sought.
The Green's function for the reverse flow operator associated with the above equation is denoted by $G_{T}$, where
\begin{equation}
k^2G_{T}+2\mathrm{i}k{\bf{M}}_0\cdot{\pmb\nabla {G}_{T}} + \nabla^2G_{T} = -\delta(\textbf{x}_p-\textbf{x}).
\label{eq:low_Mach_Number_Helmholtz_adjoint_operatror_eq}
\end{equation}
$G_{T}$ differs from $G$ and is derived subsequently in Section~\ref{sec:Green_function_adjoint_operator} using a different set of assumptions consistent with a solution in the Taylor transformed space.

Equation~(\ref{eq:low_mach_number_Helmholtz_eq}) is obtained from Eq.~(\ref{eq:convected_small_distur_Helmholtz_eq}) assuming $M^2_\infty\ll 1$ and $L_A\leq L_M$. Hence, with the same assumptions, the integral solution to Eq.~(\ref{eq:low_mach_number_Helmholtz_eq}) is derived from Eq.~(\ref{eq:Non_uniform_boundary_integral_domain}) by dropping the terms of order $M_\infty^2$ giving
\begin{equation}
\phi({\bf{x}}_p) = \int_{\partial \Omega} \left(G_{T} \frac{\partial \phi}{\partial n}- \phi\frac{\partial G_{T}}{\partial n} - 2\mathrm{i}k{\bf{M}}_0\cdot {\bf{n}} G_{T} \phi \right) dS.
\label{eq:low_Mach_number_boundary_integral_domain}
\end{equation}
Equation~(\ref{eq:low_Mach_number_boundary_integral_domain}) is written for a generic point internal to the domain $\Omega$. However, the integral becomes singular on the boundary surface~\cite{Wu2000} $\partial \Omega$. On $\partial \Omega$ the singularity is integrable in the sense of the Cauchy's principal value.

Following the procedure showed in~\ref{sec:Integral_form_along_bound_surface}, a limit approach to the boundary surface is performed. In this case, the static operator associated to Eq.~(\ref{eq:low_Mach_Number_Helmholtz_adjoint_operatror_eq}) is the Laplacian, $\nabla^2 G_{0_T}=-\delta({\bf{x}}_p-{\bf{x}})$. The static operator is the same as for the standard Helmholtz problem. Therefore, the Cauchy principal value integral, obtained in this case, is the same as for the standard Helmholtz equation~\cite{Wu2000}. By defining 
\begin{equation}
C({\bf{x}}_p)=
\left\{
\begin{array}{lr}
1 & {\bf{x}}_p\in\Omega \\
1 -&\int_{\partial \Omega} \frac{\partial G_{0_T}}{\partial n}dS \qquad {\bf{x}}_p\in\partial \Omega\\
\end{array}\right. 
\end{equation}
where $G_{0_T}=1/(4\pi R)$ and $R$ is the distance between the source and the observer, the following integral solution is obtained:
\begin{equation}
C({\bf{x}}_p)\phi({\bf{x}}_p) = \int_{\partial \Omega} \left( G_{T} \frac{\partial \phi}{\partial n}- \phi\frac{\partial G_{T}}{\partial n} - 2\mathrm{i}k{\bf{M}}_0\cdot {\bf{n}} G_{T} \phi\right) dS .
\label{eq:lowMachnumber_boundary_integral_surface_v2}
\end{equation}

\section{Green's function for weakly non-uniform mean flows}\label{sec:Green_function_adjoint_operator}

The Green's function of the reverse flow operator, Eq.~(\ref{eq:convected_small_distur_Helmholtz_adjoint_operatror_eq}), is derived by means of a variable transformation. To determine $G$, a Lorentz transformation~\cite{Chapman2000} is applied following a Taylor transformation~\cite{Taylor1979} of the fundamental reverse flow problem,
\begin{equation}
\frac{\partial^2 \hat G}{\partial t^2}-2{\bf{u}}_0\cdot{\pmb{\nabla}}\frac{\partial \hat G}{\partial t}-c_\infty^2\nabla^2\hat G+u_\infty^2\frac{\partial^2\hat G}{\partial x^2} =c_\infty^2 \delta( {\bf{x}}, t),
\label{eq:adjoint_operat_small_perturb_wave}
\end{equation}
where $\hat G = G \mathrm{e}^{\mathrm{i}\omega t}$. In the Taylor-Lorentz space, Eq.~(\ref{eq:adjoint_operat_small_perturb_wave}) reduces to the standard Helmholtz problem. Since the fundamental solution of the Helmholtz operator is well-known~\cite{Wu2000}, the Green's function in the physical space, $
G$, is retrieved by applying the inverse Taylor-Lorentz transformation to the Green's function in the transformed domain.

\par First, a Taylor transformation is applied including the non-uniform flow component ${{\bf{u}}_0'}$ to Eq.~(\ref{eq:adjoint_operat_small_perturb_wave}). The independent variables are transformed as
\begin{align}
\quad X =x,\quad  Y = y, \quad  Z = z, \quad  T = t-\frac{\Phi_0'({\bf{x}})}{c^2_\infty},
\label{eq:Taylor_Transformation_var}
\end{align}
where $\Phi_0'({\bf{x}})$ is the mean flow velocity potential corresponding to the non-uniform flow part ${\bf{u}}'_0$, ${\bf{X}}=(X,Y,Z)$ and $T$ denote the Taylor space-time, whereas $(x,y,z,t)$ denote the physical space-time. The differential operators in the Taylor space are then given using the chain rule as 
\begin{align}
\quad \frac{\partial}{\partial t}=\frac{\partial}{\partial T}, \quad {\pmb{\nabla}}= {\pmb{\nabla}}_{X} -\frac{{\bf{M}}'_0}{c_\infty} \frac{\partial}{\partial  T},
\label{eq:Taylor_Transformation_derivatives}
\end{align}
where ${\bf{M}}'_0={\bf{u}}'_0/c_\infty$.
Using the above equation, Eq.~(\ref{eq:adjoint_operat_small_perturb_wave}) can be rewritten in the Taylor space:
\begin{equation}
\begin{split}
&\frac{\partial^2 \hat G}{\partial \tilde T^2}-c_\infty^2\Bigg[\nabla_{ X}^2 \hat G-\bigg({\pmb{\nabla}}_{ X}\cdot\frac{{\bf{M}}'_0}{c_\infty }\bigg) \frac{\partial \hat G}{\partial T}-\frac{\|{\bf{M}}'_0\|^2}{c_\infty^2}\frac{\partial^2 \hat G}{\partial T^2}\Bigg] \\
&-2c_\infty{{\bf{M}}_\infty}\cdot\frac{\partial}{\partial T}\left[{\pmb{\nabla}}_{ X}\hat G-\frac{{\bf{M}}'_0}{c_\infty}\frac{\partial \hat G}{\partial T}\right]\\
&+c_\infty^2 M_\infty^2\left(\frac{\partial^2 \hat G}{\partial X^2}-2\frac{M_{0,x}'}{c_\infty}\frac{\partial^2 \hat G}{\partial X\partial T}+\frac{M'^2_{0,x}}{c^2_\infty}\frac{\partial^2 \hat G}{\partial T^2}-\frac{1}{c_\infty}\frac{\partial M_{0,x}'}{\partial X}\frac{\partial \hat G}{\partial T}\right)
= c_\infty^2 \delta( {\bf{X}},T).\\
\end{split}
\label{eq:Small_disturb_wave_equation_Taylor}
\end{equation}
Note that the Taylor transform brings the effect of the linear terms in ${\bf{M}}'_0=(M'_{0,x},M'_{0,y},M'_{0,z})$ inside the standard wave operator and introduces additional terms associated with the mean flow non-uniformities in those depending on $M_\infty$. In the above equation, consider that $M_\infty M_0'\ll M^2_\infty$ and $\nabla_{ X}  \cdot {\bf{M}}'_0$ is of order $M_0'^3$ (see Eq.~(\ref{eq:div_M0_equation})). Hence, neglecting higher order terms in $M_0'$ consistent with Eq.~(\ref{eq:convected_small_distur_wave_eq}), Eq.~(\ref{eq:Small_disturb_wave_equation_Taylor}) can be approximated by
\begin{equation}
\frac{\partial^2 \hat G}{\partial T^2}-2c_\infty{{\bf{M}}_\infty}\cdot{\pmb{\nabla}}_{ X}\frac{\partial \hat G}{\partial T} -c_\infty^2\nabla_X^2 \hat G + c_\infty^2M_\infty^2\frac{\partial^2 G}{\partial X^2} = c_\infty^2 \delta( {\bf{X}},T).
\label{eq:Small_disturb_wave_equation_Taylor_apporox}
\end{equation}
Equation~(\ref{eq:Small_disturb_wave_equation_Taylor_apporox}) is nothing but the convected wave equation in the Taylor space.

\par  Assuming a uniform mean flow aligned with the $x$-axis but opposite to the actual direction of the mean flow ${\bf{M}}_\infty$, apply a Lorentz transformation~\cite{Chapman2000,Gregory2014}. The independent variables are transformed as,
\begin{align}
\quad \tilde X = \frac{X}{\beta_\infty}, \quad {\tilde Y} = {Y}, \quad {\tilde Z} = {Z}, \quad  \tilde T = T\beta_\infty-\frac{M_\infty X}{c_\infty\beta_\infty}
\label{eq:Lorentz_Transformation_var}
\end{align}
where ${\bf{\tilde X}}=(\tilde X,\tilde Y,\tilde Z)$, $\tilde T$ denote the Taylor-Lorentz space-time and $\beta_\infty=\sqrt{1-M_\infty^2}$. The differential operators can be written as:
\begin{equation}
\begin{split}
\frac{\partial }{\partial T} =& \beta_\infty\frac{\partial}{\partial \tilde T}, \\
\frac{\partial }{\partial X} = \frac{1}{\beta_\infty}\frac{\partial}{\partial \tilde X} -\frac{M_\infty}{\beta_\infty c_\infty}\frac{\partial }{\partial \tilde T}, &\quad \quad \frac{\partial }{\partial  Y} =\frac{\partial}{\partial \tilde Y}, \quad \quad \frac{\partial }{\partial Z} =\frac{\partial}{\partial \tilde Z},\\
\end{split}
\label{eq:Lorentz_Transformation_dep_var}
\end{equation}
Equation~(\ref{eq:Lorentz_Transformation_dep_var}) is applied to Eq.~(\ref{eq:Small_disturb_wave_equation_Taylor_apporox}) giving \cite{Gregory2014}
\begin{equation}
\frac{\partial^2 \hat G}{\partial \tilde T^2}-c_\infty^2\nabla_{\tilde X}^2 \hat G= c_\infty^2 \delta( \tilde {\bf{X}},\tilde T).
\label{eq:Wave_Equation_Lorentz_Taylor}
\end{equation}
The l.h.s. of the above equation is the standard wave operator in the Taylor-Lorentz space and is independent of the mean flow.

Since a steady state problem is considered, Eq.~(\ref{eq:Wave_Equation_Lorentz_Taylor}) is rewritten in the frequency domain. The Lorentz transformation introduces a time contraction defined by the factor $\beta_\infty$ which, in turn, becomes a frequency dilation. In particular, the angular frequency in the Taylor-Lorentz space is $\tilde \omega = \omega / \beta_\infty$. On the other hand, the Taylor transformation introduces only a time delay. Therefore, Eq.~(\ref{eq:Wave_Equation_Lorentz_Taylor}) can be rewritten as,
\begin{equation}
\tilde k^2 G +\nabla_{\tilde X}^2G= -\delta( \tilde {\bf{X}}),
\label{eq:Helmholtz_Equation_Lorentz_Taylor}
\end{equation}
where $\tilde k = \tilde \omega / c_\infty $. For free field boundary conditions, the solution of Eq.~(\ref{eq:Helmholtz_Equation_Lorentz_Taylor}), given by the monopole source solution, is
\begin{equation}
G(\tilde {\bf{X}})=\frac{1}{\beta_\infty}\frac{\mathrm{e}^{-\mathrm{i}\tilde k \tilde R}}{4 \pi \tilde R } 
\label{eq:Green_function_Taylor_Lorentz}
\end{equation}
where $\tilde R=\sqrt{\tilde X^2+ \tilde Y^2 +\tilde Z^2}$. 

Equation~(\ref{eq:Green_function_Taylor_Lorentz}) is reformulated in the physical space, where a harmonic solution $G({\bf{x}},t)=G({\bf{x}})\mathrm{e}^{\mathrm{i}\omega t}$ is sought. An equivalent harmonic solution in the Taylor-Lorentz space is given by $G(\tilde {\bf{X}}, \tilde T)=G(\tilde {\bf{X}})\mathrm{e}^{\mathrm{i} \tilde\omega \tilde T}$. Applying the inverse Taylor-Lorentz transformation to the independent variables in $G(\tilde {\bf{X}}, \tilde T)$ yields
\begin{equation}
G(\tilde {\bf{X}})\mathrm{e}^{\mathrm{i} \tilde \omega \tilde T}=\bar G({\bf{x}})\mathrm{exp}\left[\mathrm{i} \frac{\omega}{\beta_\infty}\left(\beta_\infty t- \frac{M_\infty x}{c_\infty \beta_\infty} -\beta_\infty\frac{\Phi_0'({\bf{x}})}{c_\infty^2}\right)\right]= G({\bf{x}}) \mathrm{e}^{\mathrm{i} \omega t}
\label{eq:Green_function_physical_space_Taylor_Lorentz_v1}
\end{equation}
where
\begin{equation}
G({\bf{x}}) = \bar  G({\bf{x}})\mathrm{e}^{-\mathrm{i} \omega\left(\frac{M_\infty x}{c_\infty \beta^2_\infty} +\frac{\Phi_0'({\bf{x}})}{c_\infty^2}\right)}.
\label{eq:Green_function_physical_space_Taylor_Lorentz_v2}
\end{equation}
Equations~(\ref{eq:Green_function_physical_space_Taylor_Lorentz_v1}) and (\ref{eq:Green_function_physical_space_Taylor_Lorentz_v2}) give an explicit expression for the Green's function on the basis of Eq.~(\ref{eq:Green_function_Taylor_Lorentz}), i.e.,
\begin{equation}
G({\bf{x}}) =  \frac{\mathrm{exp}\left[-\mathrm{i} k\left(\frac{\sqrt{x^2+\beta^2_\infty(y^2 + z^2)}}{\beta^2_\infty}+\frac{M_\infty x}{\beta^2_\infty}+\frac{\Phi_0'({\bf{x}})}{c_\infty} \right)\right]}{4\pi\sqrt{x^2+\beta^2_\infty(y^2 + z^2)}}.
\end{equation}
The above equation is extended to a generic source position $(x_s,y_s,z_s)$, by writing
\begin{equation}
G({\bf{x}},{\bf{x}}_s) =  \frac{\mathrm{e}^{-\mathrm{i} k \sigma_{M}}}{4\pi R_M}
\label{eq:Green_function_physical_space_Taylor_Lorentz_adjoint3D}
\end{equation}
where $\sigma_{M}=[R_M+M_\infty(x-x_s)]/\beta^2_\infty + [\Phi_0'({\bf{x}})-\Phi_0'({\bf{x}}_s)]/c_\infty$ is the generalized reverse flow phase radius, extending the definition given by Garrick and Watkins~\cite{Garrick1953}, and $R_M=\sqrt{(x-x_s)^2+\beta^2_\infty[(y-y_s)^2 + (z-z_s)^2]}$ is the amplitude radius.

In the case of a uniform flow, Eq.~(\ref{eq:Green_function_physical_space_Taylor_Lorentz_adjoint3D}) is equivalent to the solution provided by Wu and Lee~\cite{Wu1994}. On the other hand, for the reverse flow Taylor Green's function $G_{T}$, the terms depending on $M^2_\infty$ are neglected in Eq.~(\ref{eq:Green_function_physical_space_Taylor_Lorentz_adjoint3D}). It follows
\begin{equation}
G_{T}({\bf{x}},{\bf{x}}_s) =  \frac{\mathrm{e}^{-\mathrm{i} k\sigma_{M_T}}}{4\pi R},
\end{equation}
where $\sigma_{M_T}=R+[\Phi_0({\bf{x}})-\Phi_0({\bf{x}}_s)]/c_\infty$, $R=\sqrt{(x-x_s)^2+(y-y_s)^2+(z-z_s)^2}$ and $\Phi_0$ denotes the total mean flow velocity potential.

\medskip 

\section{Boundary element formulation}\label{sec:variational_formulation_small_perturbation_convected_wave_equation_with_source_terms}

Consider the weakly non-uniform potential flow Helmholtz equation, Eq.~(\ref{eq:convected_small_distur_Helmholtz_eq}), with a generic distribution of harmonic sources $g({\bf{x}})$ in the domain $\Omega$
\begin{equation}
k^2\phi-2ik{\bf{M}}_0\cdot{\pmb{\nabla}} \phi + \nabla^2\phi-M_\infty^2\frac{\partial^2 \phi}{\partial x^2} = g({\bf{x}}).
\label{eq:convected_small_distur_Helmholtz_eq_with_source}
\end{equation}
Following the same procedure adopted in Section~\ref{sec:weakly_non_unif_bondary_integral_solution}, Eq.~(\ref{eq:convected_small_distur_Helmholtz_adjoint_operatror_eq}) is multiplied by $\phi$ and Eq.~(\ref{eq:convected_small_distur_Helmholtz_eq_with_source}) by $G$. The difference of these equations integrated over the domain $\Omega$ is rewritten using the divergence theorem to give
\begin{equation}
\begin{split}
\hat C({\bf{x}}_p)\phi({\bf{x}}_p) &= \int_{\Omega} G({\bf{x}}_p,{\bf{x}})  g({\bf{x}}) \ dV \\
&+ \int_{\partial \Omega}\left[ G \frac{\partial \phi}{\partial n}- \phi\frac{\partial G}{\partial n} - 2\mathrm{i}k{\bf{M}}_0\cdot {\bf{n}} G \phi \right]dS \\
&-\int_{\partial \Omega} \left[ M^2_\infty\left(G \frac{\partial \phi}{\partial x}-\phi \frac{\partial G}{\partial x}\right)n_x\right] dS ,
\label{eq:Non_uniform_boundary_integral_surface_v2_with_source}
\end{split}
\end{equation}		
where $G=G({\bf{x}}_p,{\bf{x}})$ and $\phi=\phi({\bf{x}})$ unless stated otherwise, ${\bf{x}} \in \partial \Omega$ and ${\bf{x}}_p$ is an arbitrary point either in $\Omega$ or on $\partial \Omega$. In the above equation ${\bf{M}}_0$ and ${\bf{n}}$ are given along the boundary surface $\partial \Omega$ (see Fig.~\ref{fig:geometry_Domain}). 

Equation~(\ref{eq:Non_uniform_boundary_integral_surface_v2_with_source}) can be solved numerically by means of a collocation BEM~\cite{Wu1994}. Alternatively, a variational formulation~\cite{Beriot2010} can be defined. The major downside of a variational statement is the increase in computational cost due the double integration over the boundary surface. A collocation formulation is provided below. 

First, Eq.~(\ref{eq:Non_uniform_boundary_integral_surface_v2_with_source}) is rewritten considering that 
\begin{equation}
\frac{\partial \phi}{\partial x}=\frac{\partial \phi}{\partial n}\frac{\partial n}{\partial x} +\frac{\partial \phi}{\partial \tau}\frac{\partial \tau}{\partial x} + \frac{\partial \phi}{\partial \eta}\frac{\partial \eta}{\partial x}=\frac{\partial \phi}{\partial n}n_x +\frac{\partial \phi}{\partial \tau}\tau_x+\frac{\partial \phi}{\partial \eta}\eta_x ,
\label{eq:decomposition_spacial_derivatives}
\end{equation}
where $\tau$ and $\eta$ denote the coordinates along the unit tangent vectors ${\boldsymbol{\tau}}$ and ${\boldsymbol{\eta}}$ on $\partial \Omega$, such that the normal vector to the boundary is given by ${\boldsymbol{n}}={\boldsymbol{\eta}}\times{\boldsymbol{\tau}}$. Substituting Eq.~({\ref{eq:decomposition_spacial_derivatives}}) into Eq.~(\ref{eq:Non_uniform_boundary_integral_surface_v2_with_source}) yields:
\begin{equation}
\begin{split}
&\hat C({\bf{x}}_p)\phi({\bf{x}}_p) =  \int_{\Omega} G({\bf{x}}_p,{\bf{x}})  g({\bf{x}}) \ dV  \\
&+\int_{\partial \Omega} \left[G \frac{\partial \phi}{\partial n}- \phi\frac{\partial G}{\partial n} - 2\mathrm{i}k{\bf{M}}_0\cdot {\bf{n}} G \phi \right] dS  \\
&- \int_{\partial \Omega}M^2_\infty\left[G \Bigg(\frac{\partial \phi}{\partial n}n_x +\frac{\partial \phi}{\partial \tau}\tau_x+\frac{\partial \phi}{\partial \eta}\eta_x\Bigg)-\phi \frac{\partial G}{\partial x}\right]n_x dS .
\end{split}
\label{eq:Non_uniform_boundary_integral_surface_v3}
\end{equation}
Equation~(\ref{eq:Non_uniform_boundary_integral_surface_v3}) is convenient from a computational point of view because the tangential derivative can be expressed as a sum of the shape functions multiplied by the nodal values of $\phi$ as shown by Wu and Lee \cite{Wu1994}. 

Secondly, the above equation is discretized introducing a polynomial expansion of the acoustic velocity potential and its derivatives~\cite{Wu1994}
\begin{align}
	\phi({\bf{x}})= \sum^{NDoF}_{r=1} N_r({\bf{x}}) \phi_r, && \frac{\partial \phi({\bf{x}})}{\partial n}=  \sum^{NDoF}_{r=1} N_r({\bf{x}}) \frac{\partial \phi_r}{\partial n},
	\label{eq:approx_expans_uprime}
\end{align}
\begin{align}
	\frac{\partial \phi({\bf{x}})}{\partial \eta}=  \sum^{NDoF}_{r=1} \frac{\partial N_r({\bf{x}})}{\partial \eta} \phi_r, && \frac{\partial \phi({\bf{x}})}{\partial \tau}=  \sum^{NDoF}_{r=1}  \frac{\partial N_r({\bf{x}})}{\partial \tau} \phi_r,
	\label{eq:approx_expans_duprime}
	\vspace{-5mm}
\end{align}
where $N_r({\bf{x}})$ represents the $r$-th polynomial shape function and NDoF is the total number of degrees of freedom. Hence, the discrete system of equations associated with Eq.~(\ref{eq:Non_uniform_boundary_integral_surface_v3}) can be written as
\begin{equation}
	{\mathbf{K}}^{\phi} {\pmb{\phi}}+{\bf{K}}^{u}{\frac{\partial \pmb{\phi}}{\partial n}}  = {\pmb{F}},
	\label{eq:linear_system_collocation_CHIEF}
	\vspace{-3.0mm}
\end{equation}
where
\begin{equation}
	\begin{split}
		K^{\phi}_{lm}&=	\delta_{lm}\hat C({\bf{x}}_{p,l})+\int_{\partial \Omega} 2\mathrm{i}k{\bf{M}}_0\cdot {\bf{n}} G N_m dS\\
		&+\int_{\partial \Omega}\left\{\Bigg[\frac{\partial G}{\partial n}-M^2_\infty\frac{\partial G}{\partial x} n_x\Bigg] N_m + M^2_\infty G\Bigg[\frac{\partial N_m}{\partial \tau}\tau_x+\frac{\partial N_m}{\partial \eta}\eta_x\Bigg]n_x\right\}dS\\
	\end{split}
	\label{eq:CHIEF_K_phi_matrix}
\end{equation}
\begin{align}
	K^{u}_{lm}= -\int_{\partial \Omega} G[1 - M^2_\infty n^2_x]N_mdS,
	\label{eq:CHIEF_K_u_matrix}
\end{align}
\begin{align}
F_l= \int_{\Omega} G g dV,
\label{eq:CHIEF_F_vector}
\end{align}
$G({\bf{x}}_{p,l},{\bf{x}})$ is given in Eq.~(\ref{eq:Green_function_physical_space_Taylor_Lorentz_adjoint3D}) and ${\bf{x}}_{p,l}$ denotes the p-$th$ collocation point. The vectors of the nodal degrees of freedom associated with the acoustic velocity potential and the acoustic particle velocity are denoted respectively as ${\pmb{\phi}}$ and $\partial \pmb{\phi}/\partial n$.

Similarly, for the integral formulation associated with the Taylor transformation, Eq.~(\ref{eq:Non_uniform_boundary_integral_surface_v2_with_source}) can be approximated by neglecting terms of order $M_\infty^2$. Hence, using the Green's function $G_{T}$ yields:
\begin{equation}
\begin{split}
C({\bf{x}}_p)
\phi({\bf{x}}_p) &=\int_{\Omega}G_{T}({\bf{x}}_p,{\bf{x}})  g({\bf{x}}) \ dV \\
&+ \int_{\partial \Omega} \Bigg(G_{T} \frac{\partial \phi}{\partial n}- \phi\frac{\partial G_{T}}{\partial n} - 2\mathrm{i}k{\bf{M}}_0\cdot {\bf{n}}G_{T} \phi\Bigg)dS .\\
\label{eq:Taylor_boundary_integral_surface}
\end{split}
\end{equation}
The above equation can then be discretized using Eq.~(\ref{eq:approx_expans_uprime}), but the derivation is omitted for the sake of brevity.

\section{Error estimate}\label{sec:Error_Estimate_small_perturbation_convected_wave_equation}

The weakly non-uniform potential flow wave equation, Eq.~(\ref{eq:convected_small_distur_wave_eq}), is an approximation of the full potential linearized wave equation Eq.~(\ref{eq:full_potential_acoustics}) accurate to the first order in $M_0'$. It is an exact formulation only for wave propagation on a uniform mean flow. This section presents a dimensional error analysis of the weakly non-uniform potential flow wave equation compared to the full potential linearized wave equation. If $L_A$ and $L_M$ are respectively the characteristic length scales associated with the acoustic field and with the mean flow field, Eq.~(\ref{eq:full_potential_acoustics}) can be expressed in the frequency domain as follows:
\begin{equation}
\omega^2\phi-2\mathrm{i}\omega{\bf{u}}_0\cdot {\pmb{\nabla}} \phi + c^2_\infty \nabla^2 \phi -u_\infty^2\frac{\partial^2 \phi}{\partial x^2} = \tilde E(\phi)  ,
\label{eq:linearised_helmholtz_wave_eq}
\end{equation}
where
\begin{equation}
\begin{split}
\tilde E =&{\bf{u}}'_0\cdot{\pmb{\nabla}}({\bf{u}}_0\cdot{\pmb{\nabla}}\phi)+{\bf{u}}_\infty\cdot{\pmb{\nabla}}({\bf{u}}'_0\cdot{\pmb{\nabla}}\phi)+\frac{1}{2}{\pmb{\nabla}}\phi\cdot{\pmb{\nabla}}({\bf{u}}_0\cdot{\bf{u}}_0)\\
&+(\gamma-1)(\mathrm{i}\omega \phi + {\bf{u}}_0 \cdot {\pmb{\nabla}} \phi){\pmb{\nabla}}\cdot{\bf{u}}_0+\frac{\gamma -1}{2}(\|{\bf{u}}_0\|^2-\|{\bf{u}}_\infty\|^2)\nabla^2\phi.
\end{split}
\label{eq:convected_small_distur_Helmholtz_error}
\end{equation}
In Eq.~(\ref{eq:convected_small_distur_wave_eq}) the terms on the l.h.s. of Eq.~(\ref{eq:linearised_helmholtz_wave_eq}) are retained, whereas the terms included in the error function $\tilde E$ are dropped by assuming $M_0'\ll M_\infty$.

Following Astley and Bain~\cite{Astley1986}, Eq.~(\ref{eq:convected_small_distur_Helmholtz_error}) is divided by $c^2_\infty$ and the r.h.s. terms which form $\tilde E(\phi)$ are rewritten as indicated below:
\begin{equation}
\begin{split}
&\frac{1}{c^2_\infty}{\bf{u}}'_0\cdot{\pmb{\nabla}}({\bf{u}}'_0\cdot{\pmb{\nabla}}\phi) \sim M_0^{'2}\frac{[\phi]}{L_AL_M}, \\
&\frac{1}{c^2_\infty}{\bf{u}}'_0\cdot{\pmb{\nabla}}({\bf{u}}_\infty\cdot{\pmb{\nabla}}\phi) \sim M_0^{'}M_\infty\frac{[\phi]}{L_AL_M}, \\
&\frac{1}{c^2_\infty}{\bf{u}}_\infty\cdot{\pmb{\nabla}}({\bf{u}}'_0\cdot{\pmb{\nabla}}\phi) \sim M_0^{'}M_\infty\frac{[\phi]}{L_AL_M}, \\
&\frac{1}{2 c^2_\infty}{\pmb{\nabla}}\phi\cdot{\pmb{\nabla}}({\bf{u}}_0\cdot{\bf{u}}_0) \sim M_0^{'}M_\infty\frac{[\phi]}{L_AL_M}, \\		
&\frac{1}{c^2_\infty} (\gamma-1) {\pmb{\nabla}} \cdot {\bf{u}}_0 \mathrm{i}\omega \phi  \sim M'^3_0 \frac{[\phi]}{L_AL_M}, \\
&\frac{1}{c^2_\infty} (\gamma-1) ({\pmb{\nabla}} \cdot {\bf{u}}_0) {\bf{u}}_0 \cdot {\pmb{\nabla}} \phi  \sim M'^3_0 M_\infty\frac{[\phi]}{L_AL_M}, \\
&\frac{1}{c^2_\infty}\frac{\gamma -1}{2}(\|{\bf{u}}_0\|^2-\|{\bf{u}}_\infty\|^2)\nabla^2\phi \sim M'_0 M_\infty\frac{[\phi]}{L^2_A}, \\
\end{split}
\label{eq:erorr_weakly_perturbed_potent_Helmholtz}
\end{equation}
where $[\phi]$ is the characteristic dimension of $\phi$.

To correctly bound the error brought by the solution of Eq.~(\ref{eq:Non_uniform_boundary_integral_surface_v2}), a combined Taylor-Lorentz transformation is applied to Eq.~(\ref{eq:linearised_helmholtz_wave_eq}). In the transformed space, $\phi$ and $G$ are exact solutions of the Helmholtz problem and the fundamental Helmholtz operator. Following the procedure described in Section~\ref{sec:Green_function_adjoint_operator} to obtain Eq.~(\ref{eq:Wave_Equation_Lorentz_Taylor}), but considering the actual flow direction, Eq.~(\ref{eq:linearised_helmholtz_wave_eq}) is transformed as follows:
\begin{equation}
\tilde \omega^2\phi+ c^2_\infty \nabla_{\tilde X}^2 \phi = \hat E,
\label{eq:small_disturbance_transformed_Taylor_Lorentz_errorRHS}
\end{equation}
where
\begin{equation}
\begin{split}
\hat E =& \tilde E - {\bf{u}}'_0\cdot {\bf{u}}'_0 \frac{\omega^2}{c^2_\infty} \phi + \mathrm{i} \omega ({\pmb{\nabla}}_X \cdot {\bf{u}}'_0) \phi-2\frac{\omega^2}{c_\infty^2}{\bf{u}}_\infty\cdot {\bf{u}}'_0 \phi \\
&+\mathrm{i} \omega\frac{2 u'_{0,x} u_\infty^2}{c^2_\infty} \frac{\partial \phi}{\partial X}+\omega^2 \frac{u_{0,x}^{'2} u_\infty^2 }{c^2_\infty} \phi+\frac{ u_\infty^2  }{c^2_\infty}\mathrm{i} \omega\frac{\partial u'_{0,x}}{\partial X}  \phi.
\end{split}
\label{eq:small_disturbance_transformed_Taylor_Lorentz_error}
\end{equation}
Note that the above equation is a hybrid formulation. The term $\hat E$ is given in the Taylor-Lorentz space, whereas $\tilde E$ is provided in the physical space and all of the other terms on the r.h.s. of Eq.~(\ref{eq:small_disturbance_transformed_Taylor_Lorentz_error}) are written in the Taylor space. However, Eq.~(\ref{eq:small_disturbance_transformed_Taylor_Lorentz_error}) is still suitable to estimate the order of magnitude of each term individually. The error terms in Eq.~(\ref{eq:small_disturbance_transformed_Taylor_Lorentz_error}) introduced by the Taylor-Lorentz transform can also be written, dividing again for $c_\infty^2$, as
\begin{equation}
\begin{split}
	& \frac{1}{c^4_\infty}{\bf{u}}'_0\cdot {\bf{u}}'_0 \omega^2 \phi \sim M^{'2}_0\frac{[\phi]}{L^2_A}, \\
	& \frac{1}{c^2_\infty} \mathrm{i} \omega ({\pmb{\nabla}}_X \cdot {\bf{u}}_0') \phi\sim M^{'3}_0\frac{[\phi]}{L_AL_M},  \\
	& \frac{1}{c^4_\infty} 2 {\bf{u}}'_0\cdot {\bf{u}}_\infty \omega^2  \phi\sim M'_0 M_\infty\frac{[\phi]}{L_A^2},  \\
	& \frac{1}{c^4_\infty} 2 u'_{0,x} u_\infty^2 \mathrm{i} \omega  \frac{\partial \phi}{\partial X}\sim M'_0 M_\infty^2\frac{[\phi]}{L_A^2},  \\
	& \frac{1}{c^4_\infty} u_{0,x}^{'2} u_\infty^2 \omega^2 \phi\sim M^{'2}_0 M_\infty^2\frac{[\phi]}{L_A^2},  \\
	& \frac{1}{c^4_\infty} u_\infty^2  \frac{\partial u'_{0,x}}{\partial X} \mathrm{i} \omega \phi \sim M_\infty^2 M^{'3}_0\frac{[\phi]}{L_AL_M}.  \\
\end{split}
\label{eq:erorr_weakly_perturbed_potent_Helmholtz_add_errors}
\end{equation} 
Hence, the error associated with the weakly non-uniform potential flow wave equation Eq.~(\ref{eq:small_disturbance_transformed_Taylor_Lorentz_error}) can be written as
\begin{equation}
\begin{split}
\hat E &\sim C_1 \frac{M^{'}_0M_\infty[\phi]}{L_AL_M}+C_2\frac{M'^2_0[\phi]}{L_AL_M} + C_3\frac{M^{'}_0M_\infty[\phi]}{L^2_A}+C_4\frac{M^{'2}_0[\phi]}{L_A^2} + \\
&+C_5\frac{M'^3_0[\phi]}{L_AL_M} + C_6 \frac{M'^3_0 M_\infty[\phi]}{L_AL_M}+C_7 \frac{M'_0 M_\infty^2[\phi]}{L_A^2}+C_8 \frac{M'^2_0 M_\infty^2[\phi]}{L_A^2}+C_9 \frac{M'^3_0 M_\infty^2[\phi]}{L_AL_M}.
\end{split}
\label{eq:Error_linearised_helmholtz_wave_bounded_eq}
\end{equation}
where $C_1,C_2,..,C_9$ are constants of order $1$.

From Eq.~(\ref{eq:Error_linearised_helmholtz_wave_bounded_eq}), the error $\hat E$ scales with $1/L_A=f/c_\infty$ where $f$ is the frequency. In terms of mean flow length scale, the error varies with $1/L_M$. The error therefore decreases as the mean flow becomes more uniform. The error vanishes in a uniform flow ($M'_0=0$). In other words, the accuracy of the formulation deteriorates only for wave propagation in a non-uniform region. 

The error $\varepsilon$ on the solution $\phi$ to Eq.~(\ref{eq:small_disturbance_transformed_Taylor_Lorentz_errorRHS}) is obtained by convolving $\hat E$ with the Green's function $G$, i.e.,
\begin{equation}
\varepsilon({\bf{x}}_p) = \int_{\Omega} G ({\bf{x}}_p,{\bf{x}}) \hat E({\bf{x}}) \ dV ({\bf{x}}).
\label{eq:Error_scale_transofrmations_Taylor}
\end{equation}
In the Taylor-Lorentz space $G$ varies as $1/R_M$, where $R_M$ is the amplitude radius in Eq.~(\ref{eq:Green_function_physical_space_Taylor_Lorentz_adjoint3D}). Given the geometrical length scale $D$ associated with the domain $\Omega$, the above equation can be rewritten including Eqs.~(\ref{eq:erorr_weakly_perturbed_potent_Helmholtz}) and (\ref{eq:erorr_weakly_perturbed_potent_Helmholtz_add_errors}). The error contribution provided by the second and third terms in Eq.~(\ref{eq:erorr_weakly_perturbed_potent_Helmholtz}) to Eq.~(\ref{eq:Error_scale_transofrmations_Taylor}) is given by
\begin{equation}
\varepsilon_1 \sim [\phi] M^{'}_0M_\infty \frac{D}{L_M}\frac{D^2}{L_AR_M}.
\label{eq:Error_scale_transofrmations_Taylor_dimensional_leadingerr}
\end{equation}
The error $\varepsilon_1$ increases linearly with frequency. However, if $L_AR_M$ is constant, i.e. if the amplitude radius is inversely proportional to the wavelength, the error becomes independent of frequency. 

This applies also to all the remaining terms in Eqs.~(\ref{eq:erorr_weakly_perturbed_potent_Helmholtz}) and (\ref{eq:erorr_weakly_perturbed_potent_Helmholtz_add_errors}) except for the error terms of order $1/L^2_A$. In this case, the last term of Eq.~(\ref{eq:erorr_weakly_perturbed_potent_Helmholtz}) generates an error, $\varepsilon_2$, in Eq.~(\ref{eq:Error_scale_transofrmations_Taylor}) that is given by
\begin{equation}
\varepsilon_2  \sim [\phi] M^{'}_0M_\infty D\frac{D^2}{L_A^2R_M}.
\label{eq:Error_scale_transofrmations_Taylor_dimensional_HOTerr}
\end{equation}
Note that the above equation scales quadratically with frequency. The contribution of this term is significant only at high frequency. 

Equations~(\ref{eq:Error_scale_transofrmations_Taylor_dimensional_leadingerr}) and (\ref{eq:Error_scale_transofrmations_Taylor_dimensional_HOTerr}) can also be rewritten introducing the geometrical and acoustic relative distances of the observer, i.e. $R_M/D$ and $R_M/L_A$, as
\begin{align}
& \varepsilon_1 = [\phi] M^{'}_0M_\infty \frac{D}{L_M}\left(\frac{D}{R_M}\right)^2\frac{R_M}{L_A}, \qquad \varepsilon_2 =  [\phi]  M^{'}_0M_\infty \left(\frac{D}{R_M}\right)^3\left(\frac{R_M}{L_A}\right)^2.
\end{align}
Small values of $R_M/D$ denote the geometrical near field whereas large values of $R_M/D$ are associated with the geometrical far field. Similar definitions can be given for the acoustic field on the basis of $R_M/L_A$. The above equations show that, when sound propagates in a non-uniform flow, the error decreases in the geometrical far field but increases in the acoustical far field. Contours of $\varepsilon_1$ and $\varepsilon_2$ on a logarithmic scale are shown in Figs.~\ref{fig:Figure_e1} and \ref{fig:Figure_e2}. The present error analysis can be extended to 2D problems, where $G$ scales with $1/\sqrt{kR_M}$, but it is not explicitly reported here for the sake of conciseness.
\begin{figure}[!ht]
	\subfloat[$\varepsilon_1$ (dB) \label{fig:Figure_e1}]{%
		\includegraphics[trim=1.0cm 0cm 0.8cm 0cm, clip=true, width=0.5\textwidth]{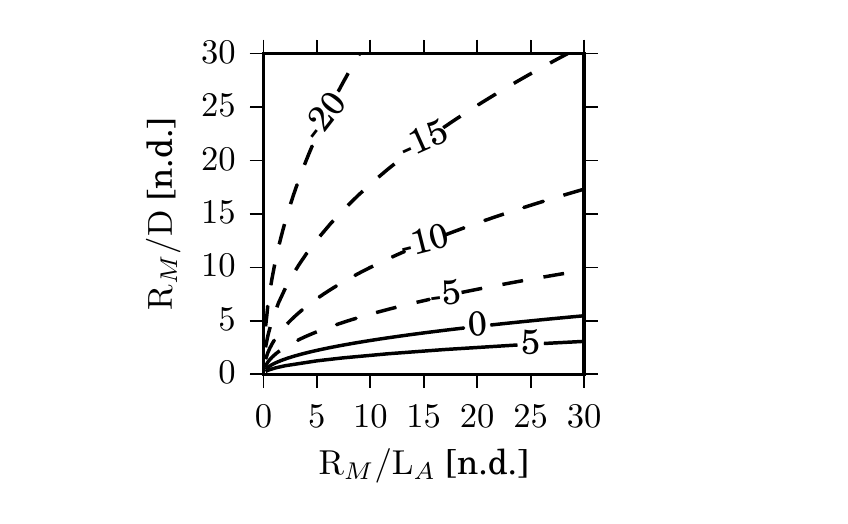}
	}
	\hfill
	\subfloat[$\varepsilon_2$ (dB)  \label{fig:Figure_e2}]{%
		\includegraphics[trim=1.0cm 0cm 0.8cm 0cm, clip=true,width=0.5\textwidth]{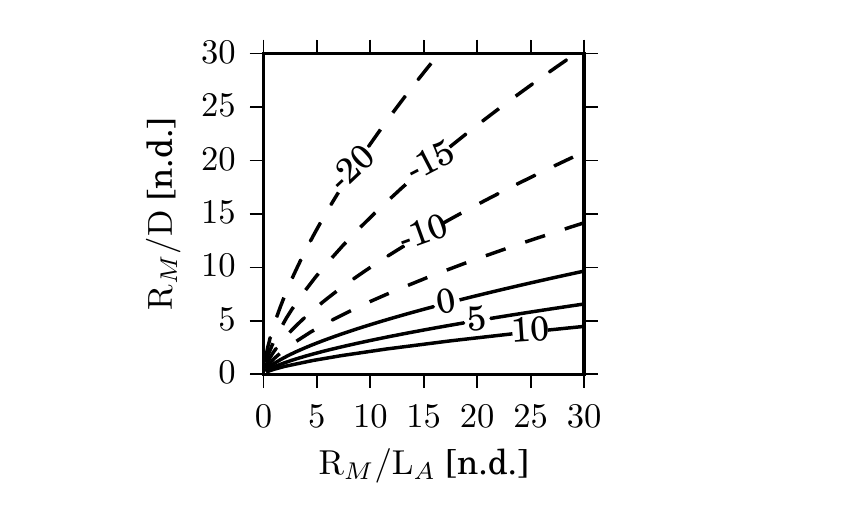}
	}
	\caption{\footnotesize Contour of the error estimates, $\varepsilon_1$ and $\varepsilon_2$, on the weakly non-uniform potential flow equation Eq.~(\ref{eq:convected_small_distur_wave_eq}) against the full potential linearized wave equation Eq.~(\ref{eq:full_potential_acoustics}).}
	\label{fig:d}
\end{figure}

The same conclusion holds for the Taylor-Helmholtz formulation. However, the error increases even when wave propagation occurs on a uniform flow. This is because the Taylor-Helmholtz equation is also an approximation of the uniform flow convected Helmholtz equation, namely the error $\hat E$ does not vanish in a uniform flow region.

\medskip

\section{Benchmark problem}\label{sec:numerical_results_BIF}

In this section, numerical examples are provided to assess the accuracy of the boundary integral solution to the weakly non-uniform potential flow Helmholtz equation Eq.~(\ref{eq:convected_small_distur_Helmholtz_eq}) compared to a reference solution of the full potential linearized Helmholtz equation based on Eq.~(\ref{eq:full_potential_acoustics}). A comparison with integral solutions to the uniform flow Helmholtz and Taylor-Helmholtz equations is also provided.

First, the integral formulations are used to solve a `wave extrapolation' problem where an `inner' solution is obtained on an arbitrary closed surface in the flow and extrapolated to the far field by using the proposed integral formulation. Second, a traditional boundary element (BE) solution is presented. The scattering of a sound field by a rigid body is computed at all points external to the scatterer up to the boundary surface. In this study, the numerical issue associated with irregular frequencies in the BE solution is overcome by avoiding the characteristic frequencies of the scatterer.

The scattering by a rigid cylinder of radius $a$ of the acoustic field generated from a monopole point source in a non-uniform mean flow is used as the basis for both the test cases. A full 2D problem is solved. The analytical solution of a potential inviscid incompressible mean flow around a cylinder without circulation is used to define the base flow and a monopole point source of unit magnitude is defined at a point ${\bf{S}}$. The problem is solved in an unbounded domain for far field Mach numbers in the range $0.0 \rightarrow 0.3$. The mean flow density $\rho_\infty=1.22$~kg m${}^{-3}$ and the speed of sound $c_\infty=340$~m s${}^{-1}$ are constant. The reference pressure for the computation of the $SPL=20\log_{10}(p_{rms}/p_{ref})$ is $p_{ref}=2\cdot10^{-5}$~Pa.

\par As a reference solution, a Lagrangian FEM with cubic element interpolation based on 30 degrees of freedom per wavelength is used to solve the full potential linearized wave equation Eq.~(\ref{eq:full_potential_acoustics}). The problem is solved in the frequency domain and a perfectly matched layer (PML) \cite{Bermudez2007} is applied to model the radiation condition. The PML is located at $r=10a$. Due to the approximation of the source model and the radiation condition, an error, measured against an analytical solution~\cite{Morris1994}, of $0.05\%$ at the outer surface of the computational domain is assessed to be provided by the FE solution in the case of quiescent media.

\medskip

\subsection{Wave extrapolation test case}

The first test case consists of a monopole point source located at $x_s=(0,-1.5a)$ as shown in Fig.~\ref{fig:geometry_FEM_IntForm}. In this case, the integral formulation is used to extrapolate the solution to the far field. First, the reference FE solution of the full potential linearized Helmholtz equation solves radiation and scattering of a monopole source by the cylinder with a non-uniform flow in the inner domain $\Omega_{in}$. Second, the reference solution is sampled on a closed control surface $\partial \Omega_{in}$, including the cylinder and the source, and radiated to the far field $\Omega_{out}$ by means of an integral formulation. In the current instance $\partial \Omega_{in}$ is located at a radial distance $r_{cs}=2a$. The maximum value of $M_0'$ on $\partial \Omega_{in}$~is~$M_\infty/4$. The order of interpolation of the integral solution on $\Omega_{in}$ is consistent with the finite element model. The FE reference solution is also used as benchmark in the far field.

\begin{figure}[h]
	\centering
	\includegraphics{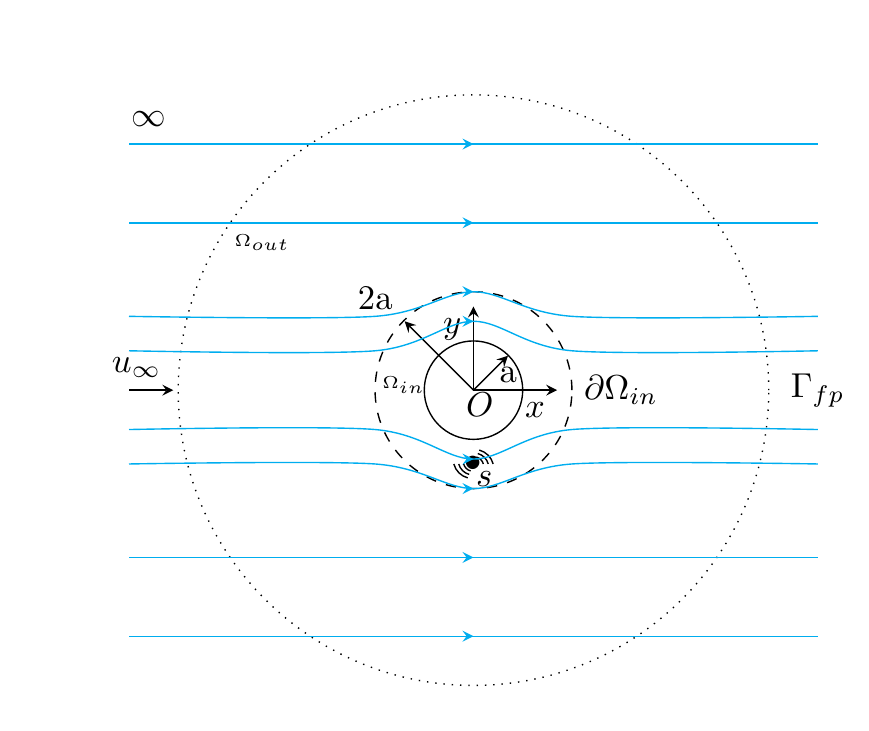} 
	\caption{\footnotesize Geometry of reference for the wave extrapolation test case. Scattering of the sound field from a monopole source on a potential mean flow by a rigid cylinder. The sound field in the outer domain $\Omega_{out}$ is extrapolated based on the field on the inner surface $\partial \Omega_{in}$.}
	\label{fig:geometry_FEM_IntForm}
\end{figure}

\begin{figure}[h]
	\centering
	\includegraphics{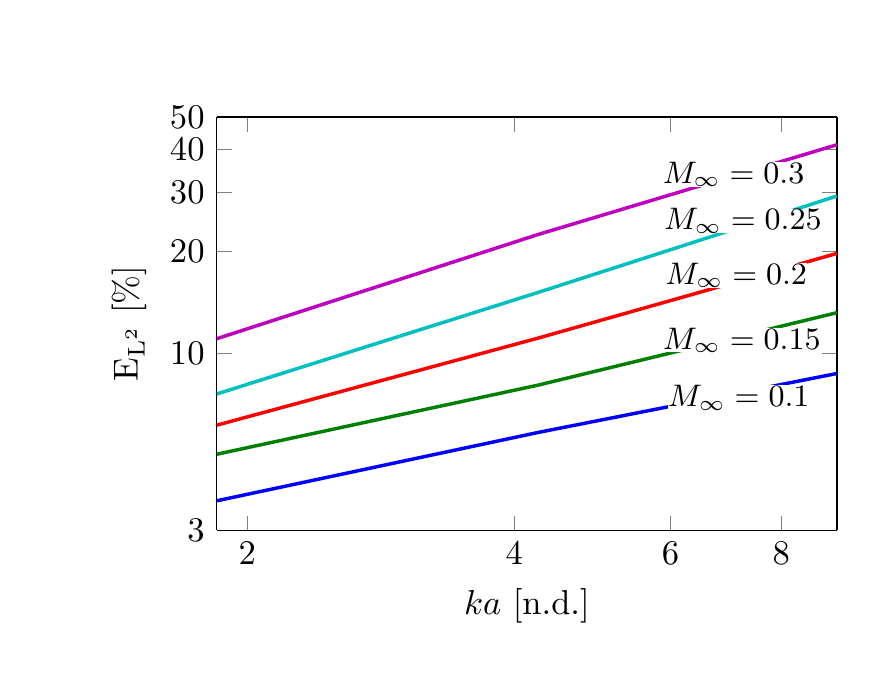} 
	\caption{\footnotesize $L^2$ error for $\phi$ along a field point circular arc with radius $r_{fp}=8a$ as a function of the non-dimensional frequency $ka$ for the wave extrapolation test case of Fig.~\ref{fig:geometry_FEM_IntForm}. The solution is based on the integral formulation for the weakly non-uniform potential flow Helmholtz equation Eq.~(\ref{eq:Non_uniform_boundary_integral_surface_v2}).}
	\label{fig:Hybrid_FEM_Integ_Taylor_Lorentz_accuracy_freq_dep} 
\end{figure}

\begin{figure}[h!]
	\centering
	\includegraphics{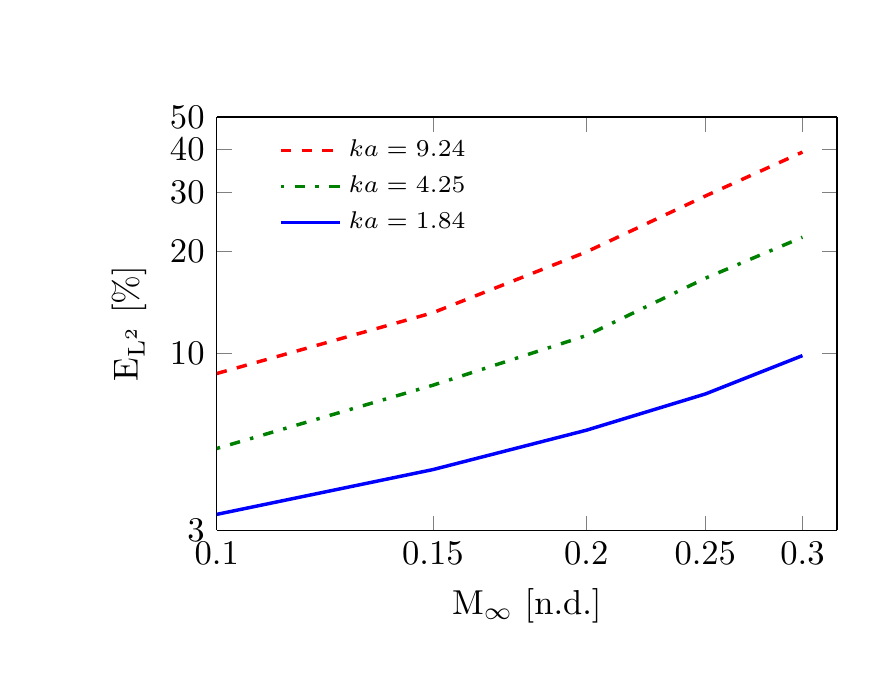} 
	\caption{\footnotesize $L^2$ error for $\phi$ along a field point circular arc with radius $r_{fp}=8a$ as a function of the mean flow Mach number $M_\infty$ for the wave extrapolation test case of Fig.~\ref{fig:geometry_FEM_IntForm}. The solution is based on the integral formulation for the weakly non-uniform potential flow Helmholtz equation Eq.~(\ref{eq:Non_uniform_boundary_integral_surface_v2}).}
	\label{fig:Hybrid_FEM_Integ_Taylor_Lorentz_accuracy_Mach}
\end{figure}

\begin{figure}[t]
	\centering
	\includegraphics{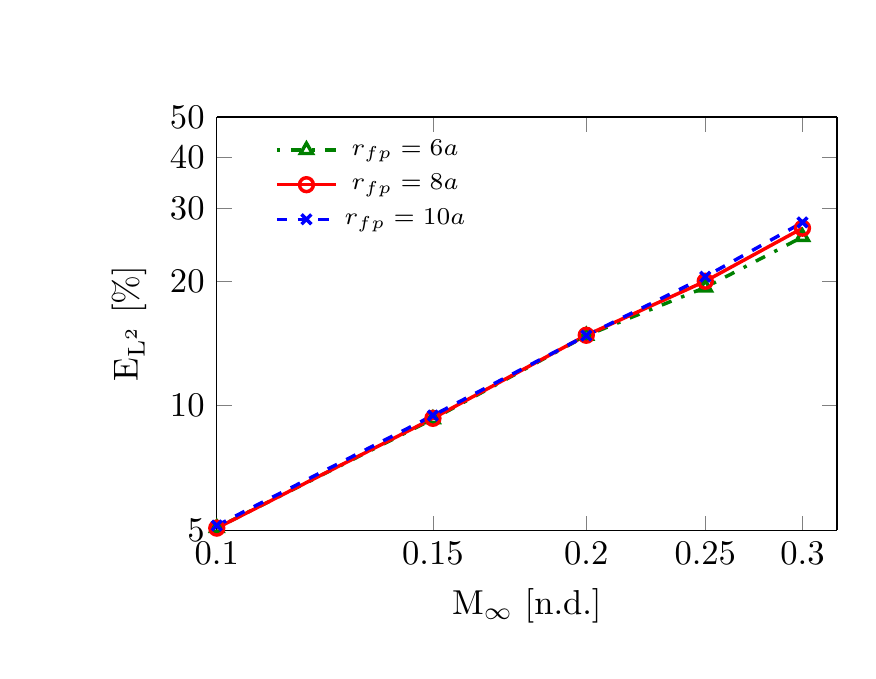} 
	\caption{\footnotesize $L^2$ error for $\phi$ at $ka=4.25$ along field point circular arcs with radii $r_{fp}=6a,8a$ and $10a$ as a function of the mean flow Mach number $M_\infty$ for the wave extrapolation test case of Fig.~\ref{fig:geometry_FEM_IntForm}. The solution is based on the integral formulation for the weakly non-uniform potential flow Helmholtz equation Eq.~(\ref{eq:Non_uniform_boundary_integral_surface_v2}).}
	\label{fig:Hybrid_FEM_Integ_Taylor_Lorentz_accuracy_dist_scale}
\end{figure}

First, the error generated by the integral solution associated with the weakly non-uniform potential flow Helmholtz equation against the reference FE solution is calculated. The $L^2$ error is defined as
\begin{equation}
E_{L^2}= 100\times\sqrt{\frac{\int_{\Gamma_{fp}}\| \phi({\bf{x}}) - \phi({\bf{x}})_{ref}\|^2dS}{\int_{\Gamma_{fp}}\| \phi_{ref}({\bf{x}})\|^2dS}}
\label{eq:L2_norm_error_def}
\end{equation} 
where $\Gamma_{fp}$ denotes a closed circular arc of field points (see Fig.~\ref{fig:geometry_FEM_IntForm}). The $L^2$ error for the acoustic velocity potential is shown in Fig.~\ref{fig:Hybrid_FEM_Integ_Taylor_Lorentz_accuracy_freq_dep} plotted against the Helmholtz number $ka$ on a circular arc of field points with radius $r_{fp}=8a$. Note that the error is computed considering the physical distance of the observer $R$ in lieu of the amplitude radius $R_M$ since for $M_\infty\leq0.3$ the maximum difference between $R$ and $R_M$ is about $5\%$ of R. The variation of the $L^2$ error for $\phi$ against $M_\infty$ is shown in Fig.~\ref{fig:Hybrid_FEM_Integ_Taylor_Lorentz_accuracy_Mach} whereas the sensitivity of the $L^2$ error on the distance to the observer is shown in Fig.~\ref{fig:Hybrid_FEM_Integ_Taylor_Lorentz_accuracy_dist_scale}. In the latter, the error is computed on circular arcs of radii $r_{fp}$ equal to $6a$, $8a$ and $10a$ for $ka=4.25$. Note that the mean flow is almost uniform at these distances from the cylinder. To further assess the accuracy of the integral formulation for the weakly non-uniform potential flow Helmholtz equation, the computation of the $L^2$ error for $\phi$ against $M_\infty$ is also performed for the formulation of Wu and Lee~\cite{Wu1994} and the Taylor-Helmholtz equation Eq.~(\ref{eq:Taylor_boundary_integral_surface}). This comparison is given in Figs.~\ref{fig:Figure_7a}, \ref{fig:Figure_7b} and \ref{fig:Figure_7c} computing the error along an arc of field points with radius $r_{fp}=8a$ for different values of the the non-dimensional frequency $ka$.

%\begin{figure}[t]
%	\centering
%	\includegraphics{Figure/Figure_7.eps} 
%	\caption{\footnotesize $L^2$ error for $\phi$ along a field point circular arc with radius $r_{fp}=8a$ as a function of the mean flow Mach number $M_\infty$ for the wave extrapolation test case of Fig.~\ref{fig:geometry_FEM_IntForm} at  $ka=1.84$ (solid), $ka=4.25$ (dash) and $ka=9.24$ (dash-dot). The solutions are based on the weakly non-uniform potential flow Helmholtz integral formulation Eq.~(\ref{eq:Non_uniform_boundary_integral_surface_v2}) ($\triangle$), the uniform flow Helmholtz integral formulation\cite{Wu1994} ($\circ$) and the Taylor-Helmholtz integral formulation Eq.~(\ref{eq:lowMachnumber_boundary_integral_surface_v2}) ($\times$).}
%	\label{fig:BEM_Integ_Taylor_Lorentz_accuracy_L2}
%\end{figure}

\begin{figure}[!ht]
	\subfloat[$ka=1.94$\label{fig:Figure_7a}]{%
		\includegraphics[trim=1.0cm 0cm 0.3cm 0cm, clip=true, width=0.32\textwidth]{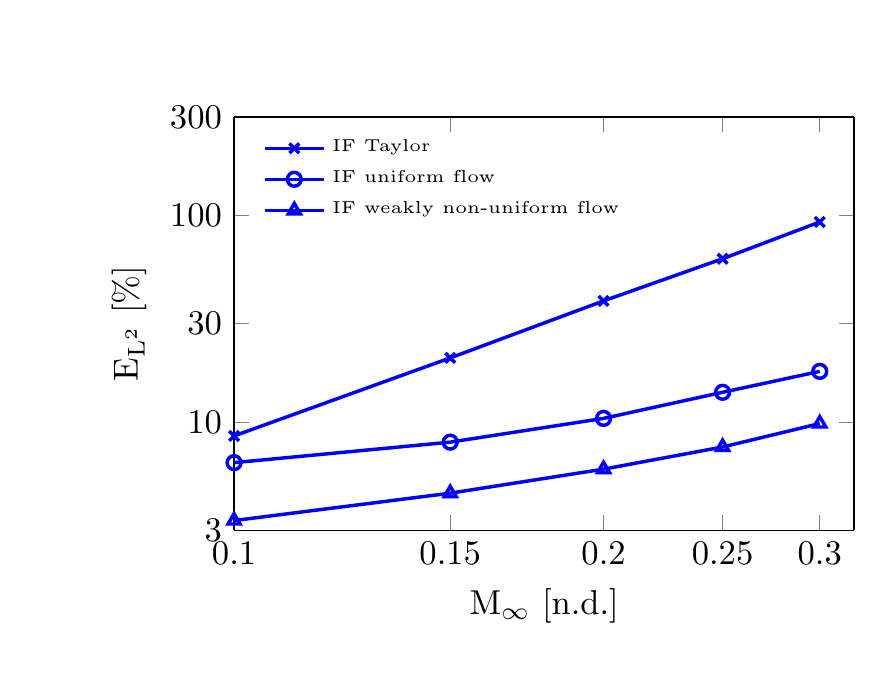}
	}
	\hfill
	\subfloat[$ka=4.25$\label{fig:Figure_7b}]{%
		\includegraphics[trim=1.0cm 0cm 0.3cm 0cm, clip=true,width=0.32\textwidth]{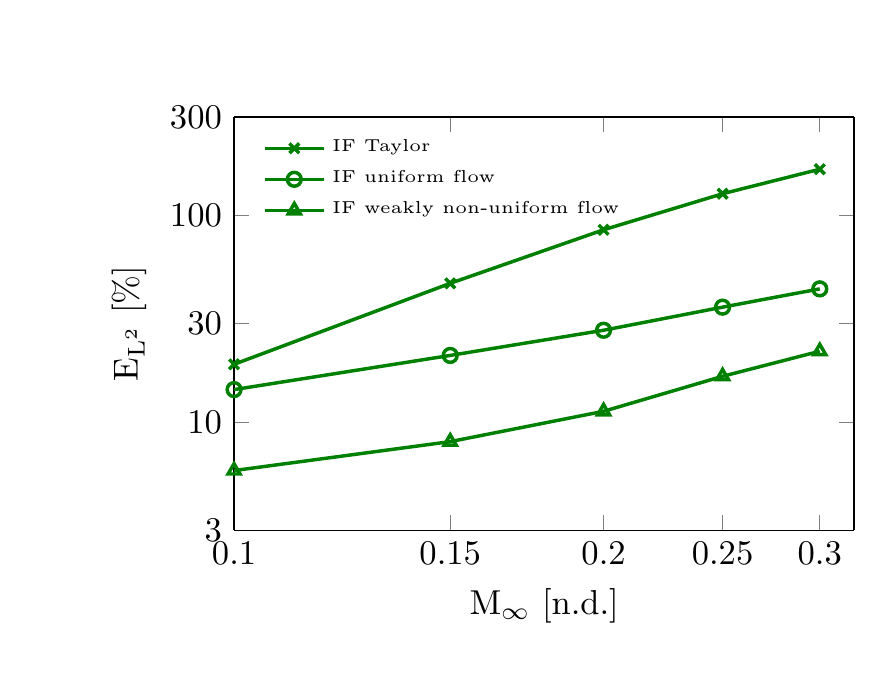}
	}
	\hfill
	\subfloat[$ka=9.24$\label{fig:Figure_7c}]{%
		\includegraphics[trim=1.0cm 0cm 0.3cm 0cm, clip=true,width=0.32\textwidth]{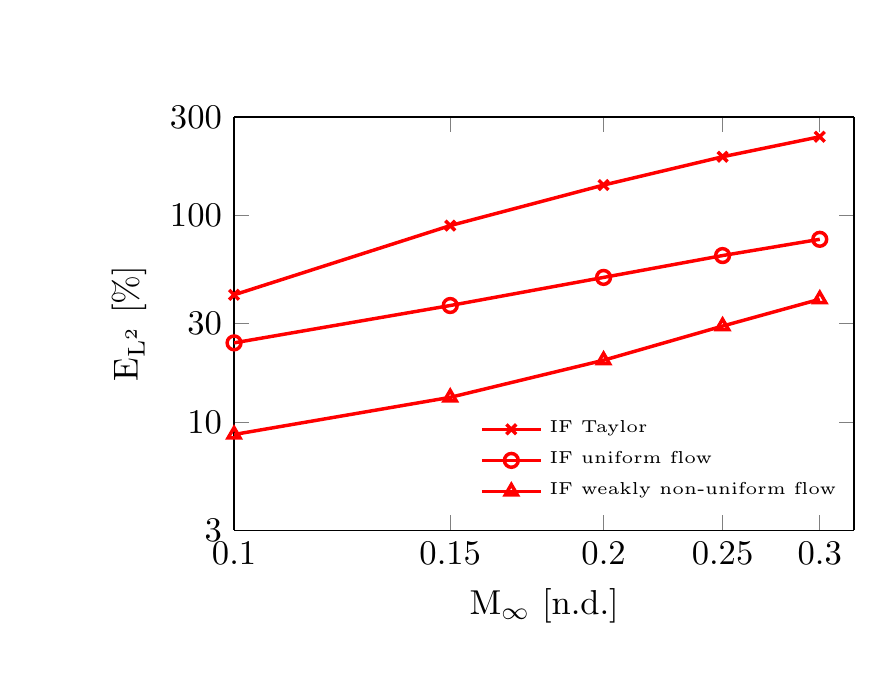}
	}
	\caption{\footnotesize $L^2$ error for $\phi$ along a field point circular arc with radius $r_{fp}=8a$ as a function of the mean flow Mach number $M_\infty$ for the wave extrapolation test case of Fig.~\ref{fig:geometry_FEM_IntForm} at  $ka=1.84$, $4.25$ and $9.24$. The solutions are based on the weakly non-uniform potential flow Helmholtz integral formulation Eq.~(\ref{eq:Non_uniform_boundary_integral_surface_v2}) ($\triangle$), the uniform flow Helmholtz integral formulation~\cite{Wu1994} ($\circ$) and the Taylor-Helmholtz integral formulation Eq.~(\ref{eq:lowMachnumber_boundary_integral_surface_v2}) ($\times$).}
	\label{fig:BEM_Integ_Taylor_Lorentz_accuracy_L2}
\end{figure}

\begin{figure}[t!]
	\centering
	\includegraphics{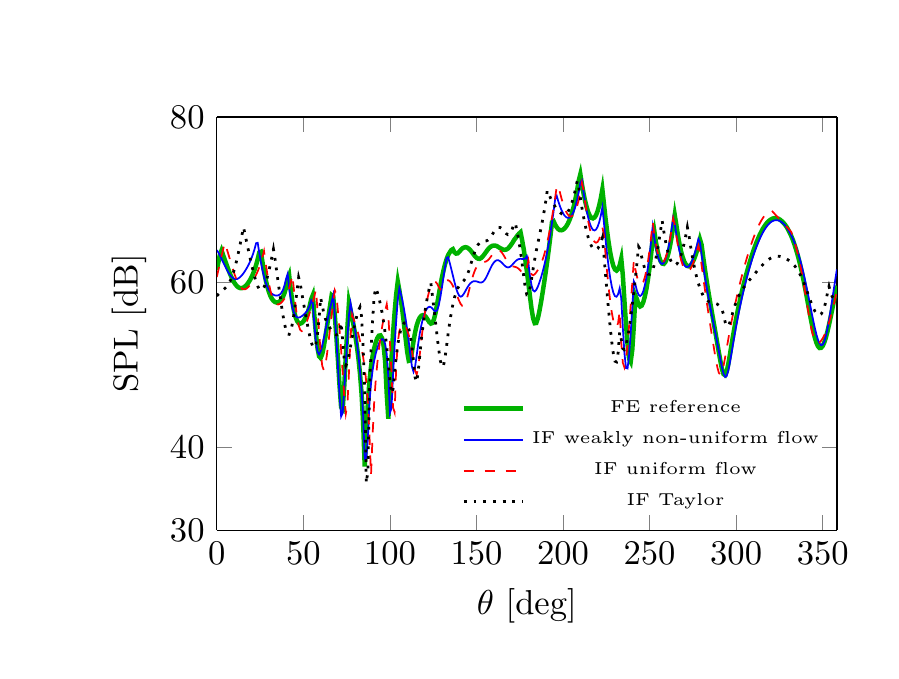} 
	\caption{\footnotesize Acoustic pressure directivity along a field point circular arc with radius $r_{fp}=8a$ for $ka=9.24$ and $M_\infty=0.3$. The wave extrapolation solution (see Fig.~\ref{fig:geometry_FEM_IntForm}) is based on the integral formulations Eq.~(\ref{eq:Non_uniform_boundary_integral_surface_v2}) (IF weakly non-uniform), Wu and Lee~\cite{Wu1994} (IF uniform flow) and Eq.~(\ref{eq:lowMachnumber_boundary_integral_surface_v2}) (IF Taylor).}
	\label{fig:directivity_wave_extrapolation_k9_M03}
\end{figure}

\par A description of the local accuracy provided by the three integral formulations is presented in Fig.~\ref{fig:directivity_wave_extrapolation_k9_M03}, which illustrates the pressure directivity at $r_{fp}=8a$ for $ka=9.24$ and $M_\infty=0.3$ plotted against the angle $\theta$, where $\theta=0$ corresponds to the $x$-axis and $\theta$ is measured counterclockwise. The acoustic pressure is given by
\begin{equation}
p=-\rho_0(\mathrm{i}\omega\phi+{\bf{u}}_0\cdot{\pmb{\nabla}}\phi).
\end{equation}
Again, the FE solution of the full potential linearized Helmholtz equation is used as a reference result. The solution provided by the Taylor-Helmholtz equation introduces local errors of up to 10 dB. On the other hand, the error given by the weakly non-uniform potential flow Helmholtz solution and the related uniform flow approximation is limited to 5 dB. However, the weakly non-uniform solution improves the results given by the uniform flow Helmholtz equation in the shielded area [$\theta=60^\circ-120^\circ$], where the mean flow is not aligned with the uniform stream.

The accuracy of the proposed integral formulation Eq.~(\ref{eq:Non_uniform_boundary_integral_surface_v2}) decreases linearly with frequency (see Fig.~\ref{fig:Hybrid_FEM_Integ_Taylor_Lorentz_accuracy_freq_dep}) and Mach number $M_\infty$ (see Fig.~\ref{fig:Hybrid_FEM_Integ_Taylor_Lorentz_accuracy_Mach}), whereas the error is almost constant when wave propagation occurs on a uniform mean flow (see Fig.~\ref{fig:Hybrid_FEM_Integ_Taylor_Lorentz_accuracy_dist_scale}) since the integral formulation associated with the weakly non-uniform potential flow Helmholtz equation is exact for a uniform base flow. These results validate the error analysis of Section~\ref{sec:Error_Estimate_small_perturbation_convected_wave_equation}. 

Moreover, the integral formulation based on the weakly non-uniform potential flow Helmholtz equation provides the best accuracy independently of frequency and Mach number compared to the solution obtained using the integral formulations for the uniform flow Helmholtz and the Taylor-Helmhotlz equations (see Fig.~\ref{fig:BEM_Integ_Taylor_Lorentz_accuracy_L2}). The non-uniform flow effects in the weakly non-uniform formulation improve the prediction made compared to the uniform flow Helmholtz equation. The Taylor formulation however performs poorly and is less accurate than the uniform flow formulation which assumes uniform flows at all points. The problem with the Taylor equation is that the error grows even in the uniform flow region whereas the uniform and the weakly non-uniform flow models are exact once this domain is reached.

\begin{figure}[h!]
	\centering
	\includegraphics[width=0.8\textwidth]{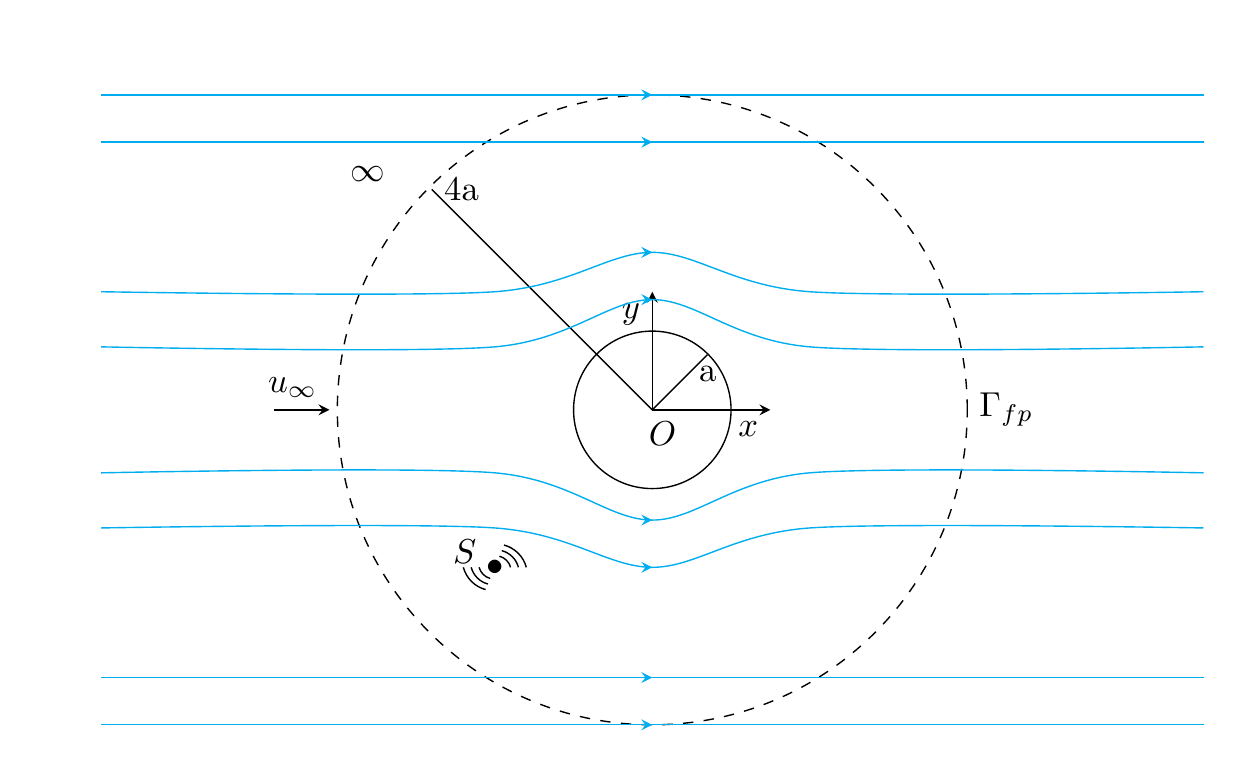} 
	\caption{\footnotesize Reference domain for the BE solution showing the non-uniform mean flow streamlines. Scattering of the sound field from a monopole source $S$ by a rigid cylinder in an unbounded domain with non-uniform flow.}
	\label{fig:geometry_BEM}
\end{figure}

\subsection{Boundary element test case}

In this section, the accuracy of the boundary integral formulation for the weakly non-uniform potential flow Helmholtz equation is assessed by considering a direct BE solution of Eq.~(\ref{eq:Non_uniform_boundary_integral_surface_v3}), solving for the surface acoustic velocity potential on the cylinder due to a sound field from a point source (see Fig.~\ref{fig:geometry_BEM}). The integral formulation is used to represent wave radiation and scattering up to the boundary surface, where the non-uniform mean flow component $M_0'$ is of the same order of magnitude as $M_\infty$. The cylinder with radius $a$ is centered at the origin of the reference frame and a monopole source is located at $x_s=(-2a,-2a)$ as shown in Fig.~\ref{fig:geometry_BEM}. The error in the solution is calculated against the reference FE solution. A cubic Lagrangian BE interpolation with 10~DoFs/$\lambda$ is used to solve the integral equations.

\begin{figure}[t]
	\centering
	\includegraphics{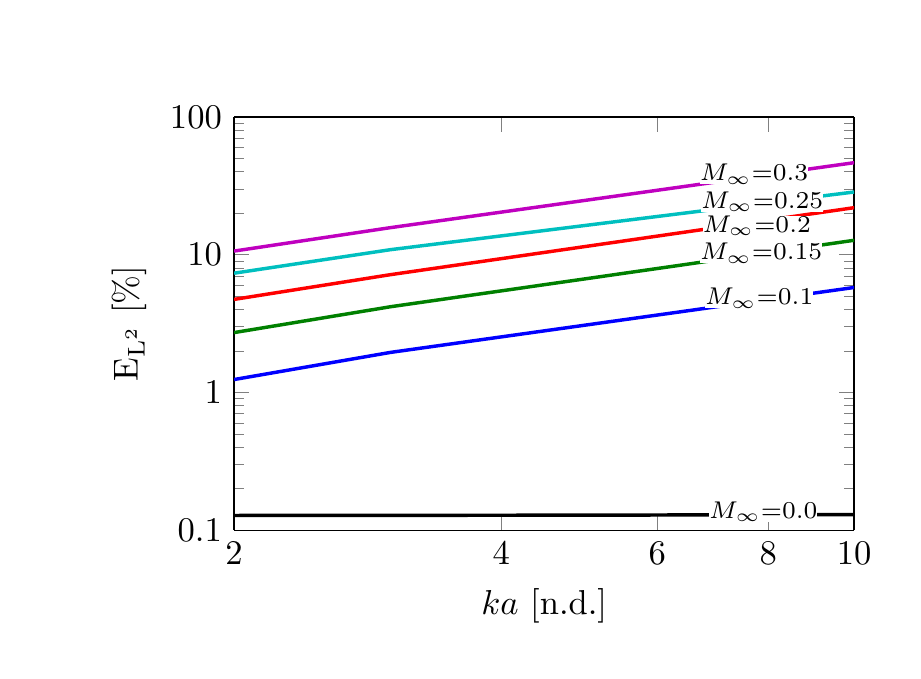} 
	\caption{\footnotesize $L^2$ error for $\phi$ along a field point circular arc with radius $r_{fp}=4a$ for the BE solution of the problem in Fig.~\ref{fig:geometry_BEM} as a function of the non-dimensional frequency $ka$. The solution is based on Eq.~(\ref{eq:Non_uniform_boundary_integral_surface_v3}).}
	\label{fig:BEM_Non_Uni_indErr_freqSens}
\end{figure}

\begin{figure}[bh!]
	\centering
	\includegraphics{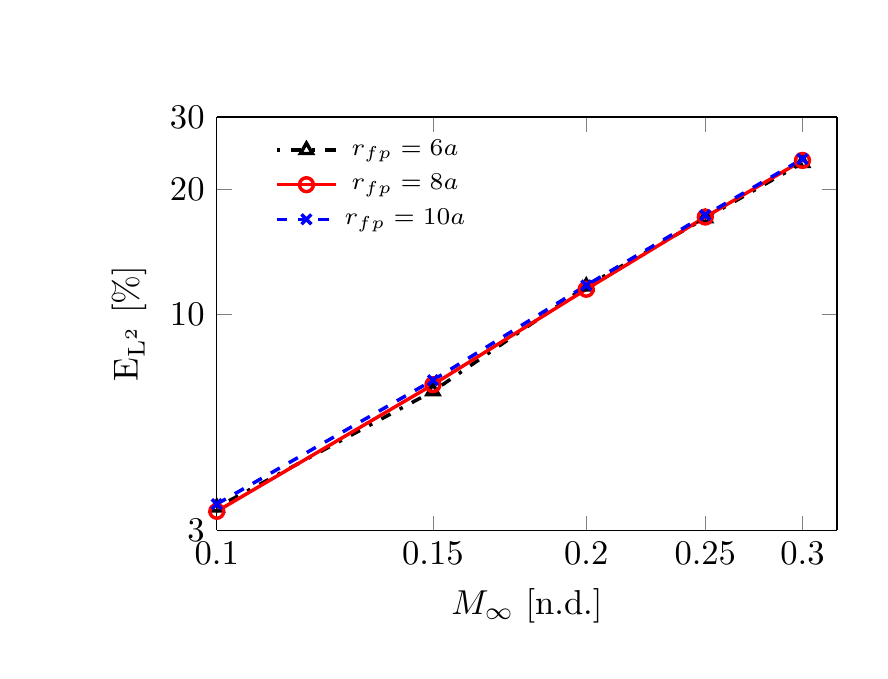} 
	\caption{\footnotesize $L^2$ error for $\phi$ at a non-dimensional frequency $ka=5$ along field point circular arcs with radii $r_{fp}=6a,8a$ and $10a$ for the BE solution of the problem in Fig.~\ref{fig:geometry_BEM} as a function of the mean flow Mach number $M_\infty$. The solution is based on Eq.~(\ref{eq:Non_uniform_boundary_integral_surface_v3}).}
	\label{fig:BEM_Non_Uni_indErr_Unif_flow}
\end{figure}

\par The dependency of the $L^2$ error for $\phi$ against the non-dimensional frequency at a circular arc of field points with radius $r_{fp}=4a$ is illustrated in Fig.~\ref{fig:BEM_Non_Uni_indErr_freqSens}. Again, note that the error is computed considering the physical distance of the observer, denoted previously as $R$, in lieu of $R_M$ since for $M_\infty\leq0.3$ the maximum difference between $R$ and $R_M$ is about $5\%$ of $R$. Figure~\ref{fig:BEM_Non_Uni_indErr_Unif_flow} shows the sensitivity of the $L^2$ error for $\phi$ to the distance of the observer sampling the error on circular arcs where the flow is almost uniform, for a non-dimensional frequency $ka=5$. 

A comparison of the accuracy of the different integral solutions is also shown in Figs.~\ref{fig:Figure_12a}, \ref{fig:Figure_12b} and \ref{fig:Figure_12c} where the $L^2$ error for $\phi$ is computed at an arc of field points with radius $r_{fp}=4a$ for the weakly non-uniform potential flow Helmholtz equation, the uniform flow Helmholtz equation~\cite{Wu1994} and the Taylor-Helmholtz equation. Figures \ref{fig:Figure_13a}, \ref{fig:Figure_13b} and \ref{fig:Figure_13c} show contours of the real part of $\phi$ over the solution domain for $ka=10$ and $M_\infty=0.3$. The weakly non-uniform potential flow formulation (Fig.~\ref{fig:Figure_13a}) clearly approximates the full potential FE solution (Fig.~\ref{fig:Figure_13b}) more accurately than the uniform flow Helmholtz solution (Fig.~\ref{fig:Figure_13c}). This is more evident if the sound source is in a region where the mean flow is strongly non-uniform, as shown in Fig.~\ref{fig:dummy_test2}, where the same problem as in Fig.~\ref{fig:geometry_BEM} is solved locating the monopole point source at ${\bf{x}}_s=(-1.3a,-0.5a)$.

%\begin{figure}[t!]
%	\centering
%	\includegraphics{Figure/Figure_12.eps} 
%	\caption{\footnotesize $L^2$ error for $\phi$ along a field point circular arc with radius $r_{fp}=4a$ as a function of the mean flow Mach number $M_\infty$ at non-dimensional frequencies $ka=2$ (solid), $ka=5$ (dash-dot) and $ka=10$ (dash) for the problem in Fig.~\ref{fig:geometry_BEM}. The solutions are based on the weakly non-uniform potential flow Helmholtz equation Eq.~(\ref{eq:Non_uniform_boundary_integral_surface_v3}) ($\triangle$), the uniform flow Helmholtz equation\cite{Wu1994} ($\circ$) and the Taylor-Helmholtz equation Eq.~(\ref{eq:Taylor_boundary_integral_surface}) ($\times$).}
%	\label{fig:BEM_Integ_Taylor_Lorentz_accuracy_L2err}
%\end{figure}

\begin{figure}[!ht]
	\subfloat[$ka=2$\label{fig:Figure_12a}]{%
		\includegraphics[trim=1.0cm 0cm 0.3cm 0cm, clip=true, width=0.32\textwidth]{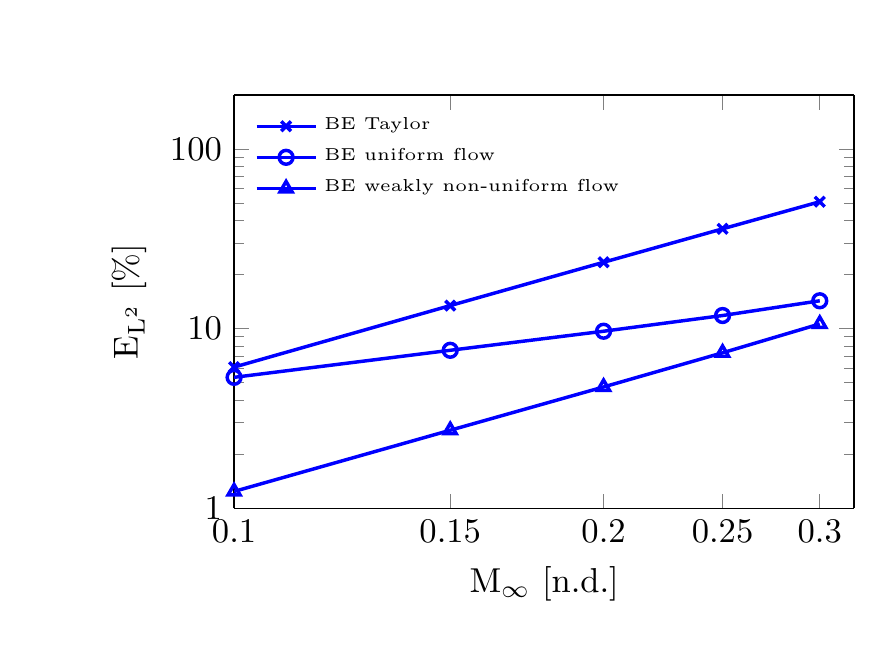}
	}
	\hfill
	\subfloat[$ka=5$\label{fig:Figure_12b}]{%
		\includegraphics[trim=1.0cm 0cm 0.3cm 0cm, clip=true,width=0.32\textwidth]{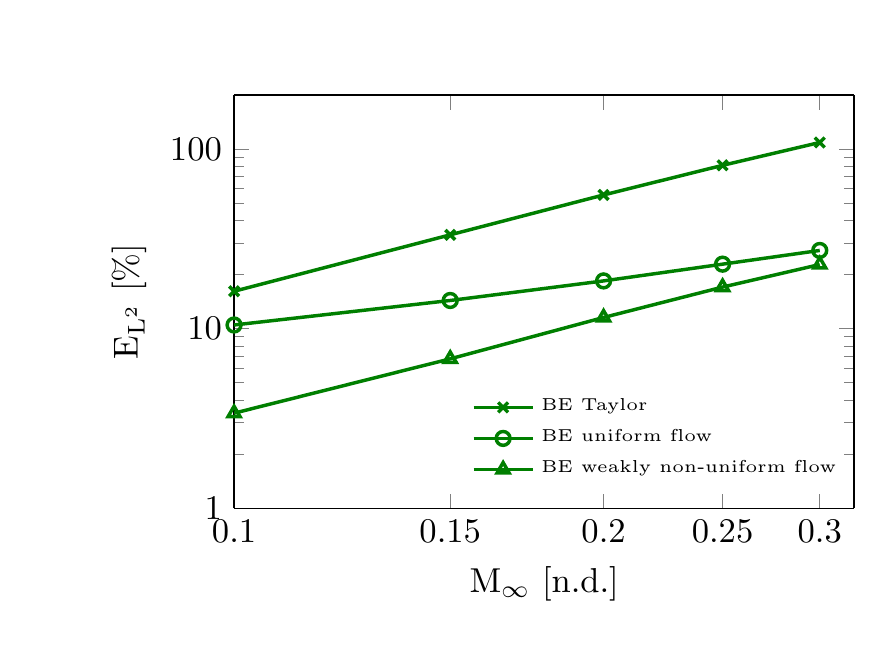}
	}
	\hfill
	\subfloat[$ka=10$\label{fig:Figure_12c}]{%
		\includegraphics[trim=1.0cm 0cm 0.3cm 0cm, clip=true,width=0.32\textwidth]{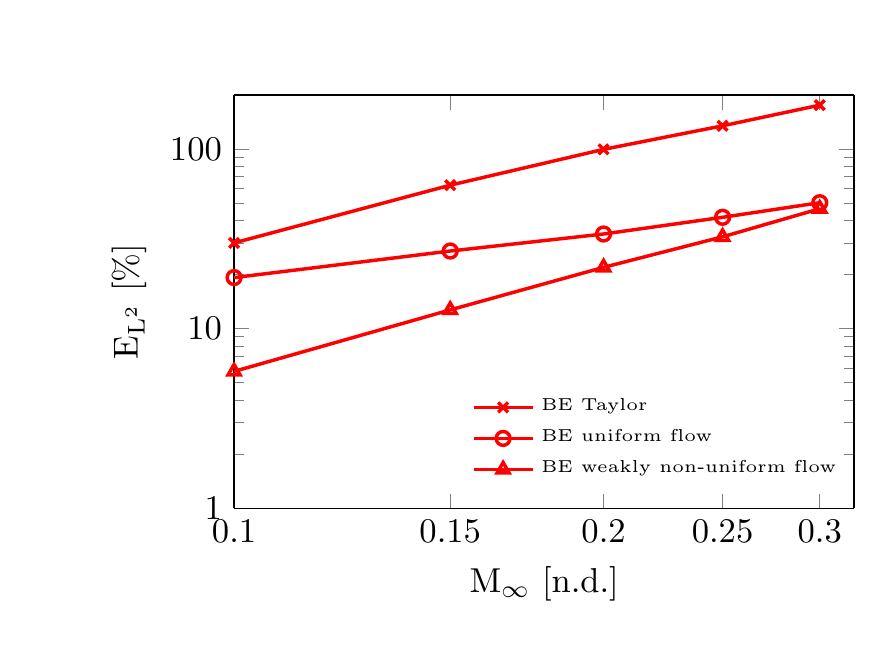}
	}
\caption{\footnotesize $L^2$ error for $\phi$ along a field point circular arc with radius $r_{fp}=4a$ as a function of the mean flow Mach number $M_\infty$ at non-dimensional frequencies $ka=2$, $5$ and $10$ for the problem in Fig.~\ref{fig:geometry_BEM}. The solutions are based on the weakly non-uniform potential flow Helmholtz equation Eq.~(\ref{eq:Non_uniform_boundary_integral_surface_v3}) ($\triangle$), the uniform flow Helmholtz equation~\cite{Wu1994} ($\circ$) and the Taylor-Helmholtz equation Eq.~(\ref{eq:Taylor_boundary_integral_surface}) ($\times$).}
\label{fig:BEM_Integ_Taylor_Lorentz_accuracy_L2err}
\end{figure}	

\begin{figure}[!ht]
	\subfloat[\label{fig:Figure_13a}]{%
		\includegraphics[trim=1.3cm 0cm 0.3cm 0cm, clip=true, width=0.32\textwidth]{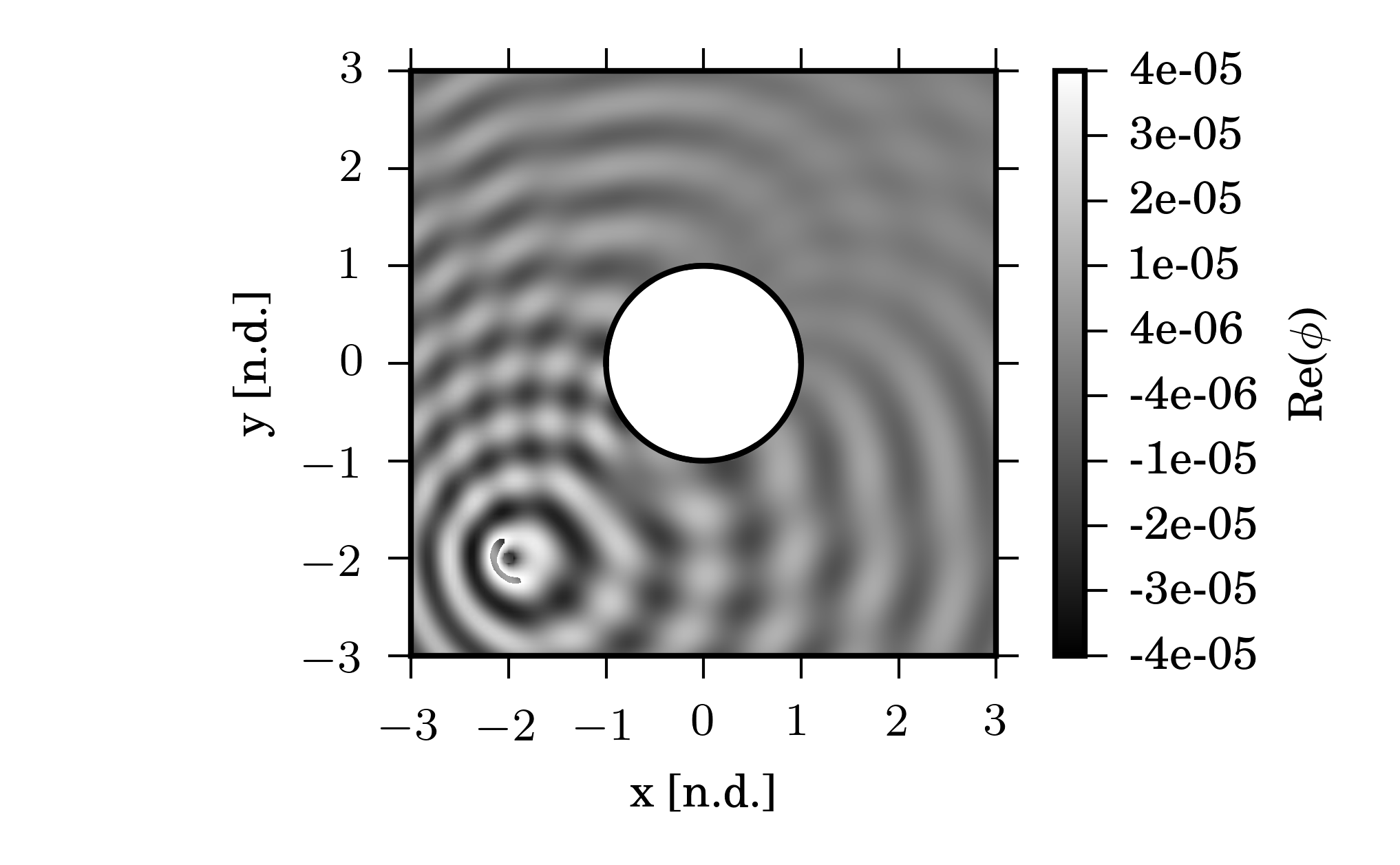}
	}
	\hfill
	\subfloat[\label{fig:Figure_13b}]{%
		\includegraphics[trim=1.3cm 0cm 0.3cm 0cm, clip=true,width=0.32\textwidth]{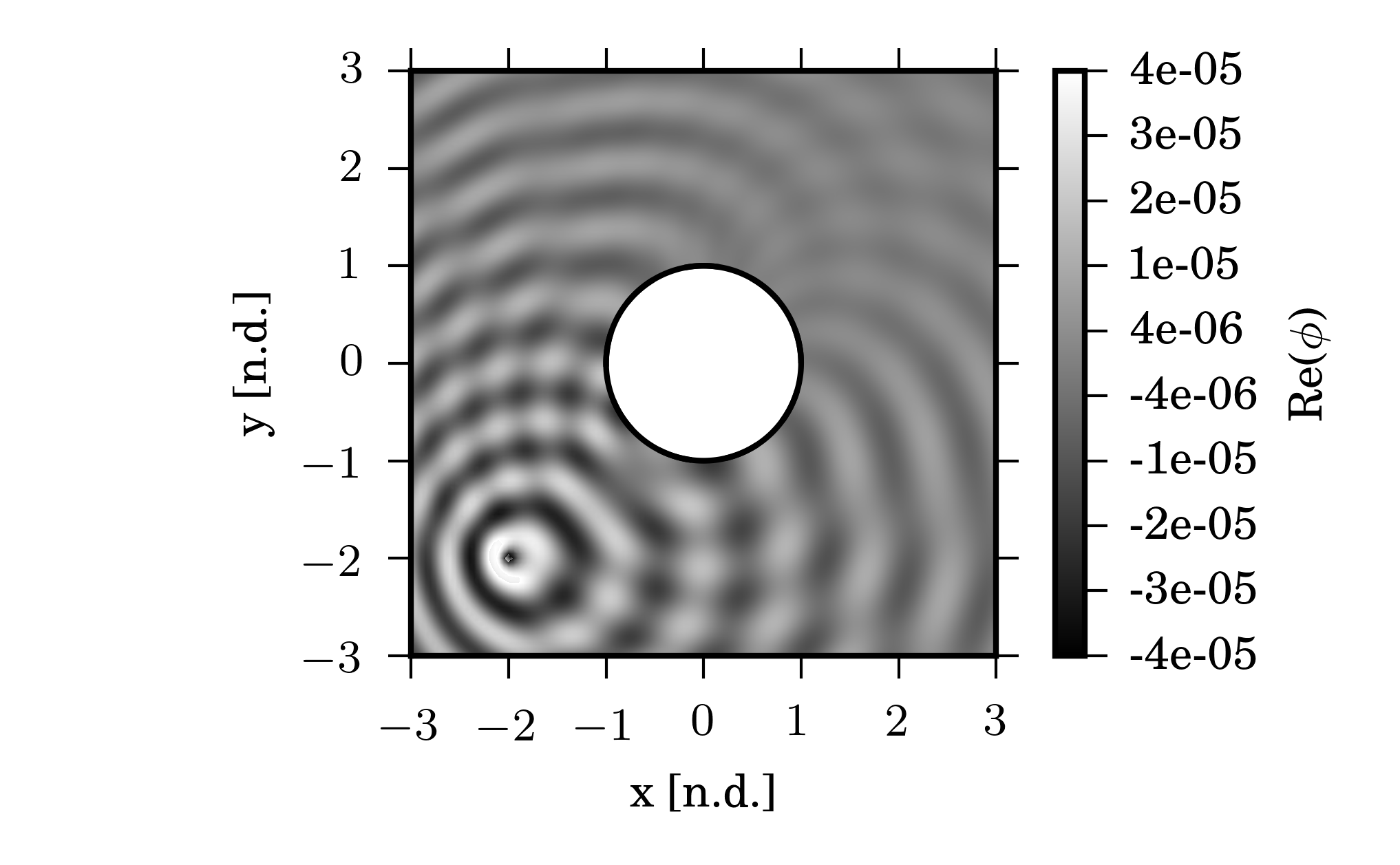}
	}
	\hfill
	\subfloat[\label{fig:Figure_13c}]{%
		\includegraphics[trim=1.3cm 0cm 0.3cm 0cm, clip=true,width=0.32\textwidth]{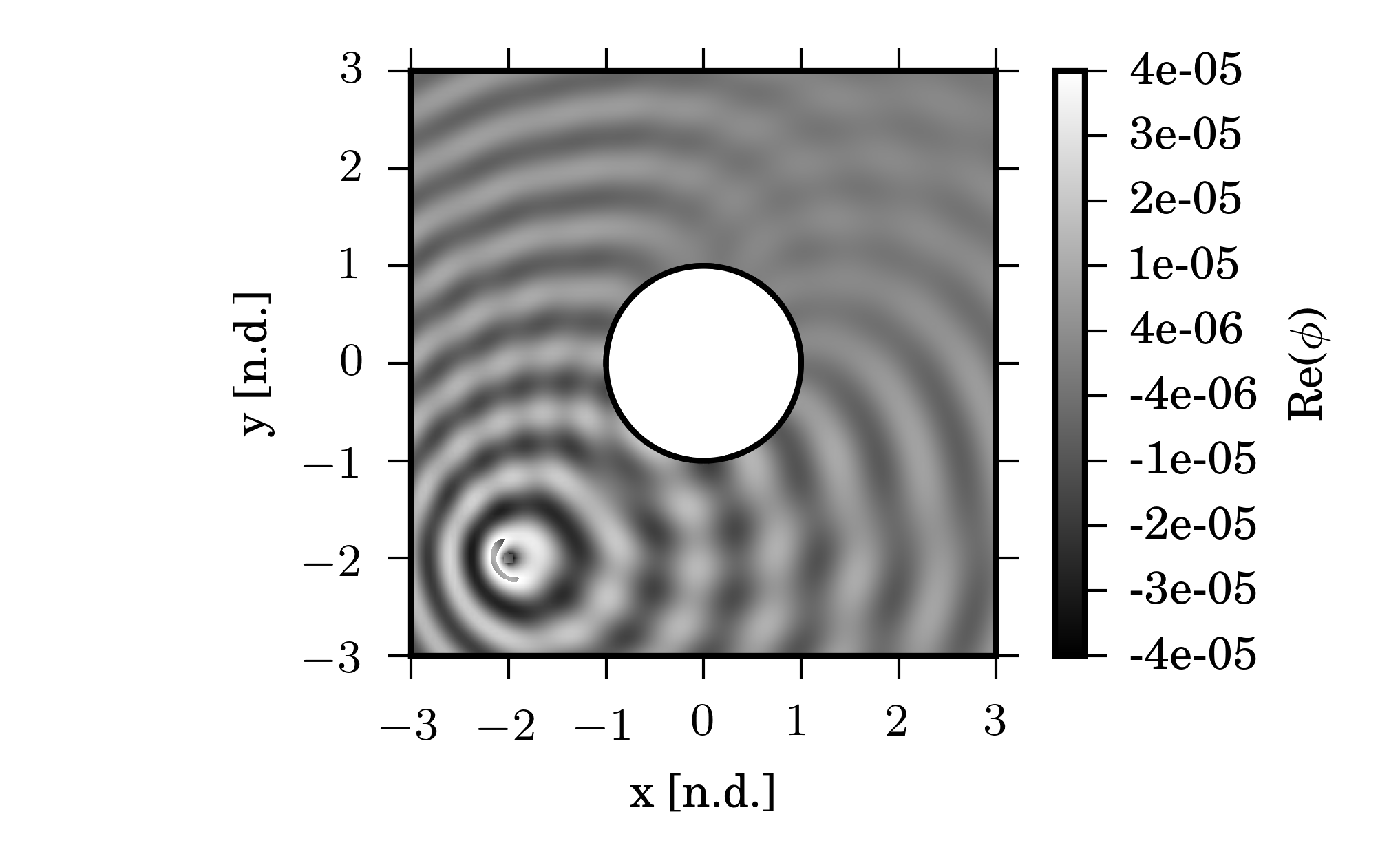}
	}
	\caption{\footnotesize  Real part of the acoustic velocity potential, $Re(\phi)$, at a non-dimensional frequency $ka=10$ for $M_\infty=0.3$. The BE solution of the weakly non-uniform potential flow Helmholtz equation (a), the FE solution of the full potential linearized Helmholtz equation (b) and the BE solution of the uniform flow Helmholtz equation~\cite{Wu1994} (c) are shown for the problem in Fig.~\ref{fig:geometry_BEM}.}
	\label{fig:dummy}
\end{figure}

\begin{figure}[!ht]
	\subfloat[\label{fig:Figure_14a}]{%
		\includegraphics[trim=1.3cm 0cm 0.3cm 0cm, clip=true, width=0.32\textwidth]{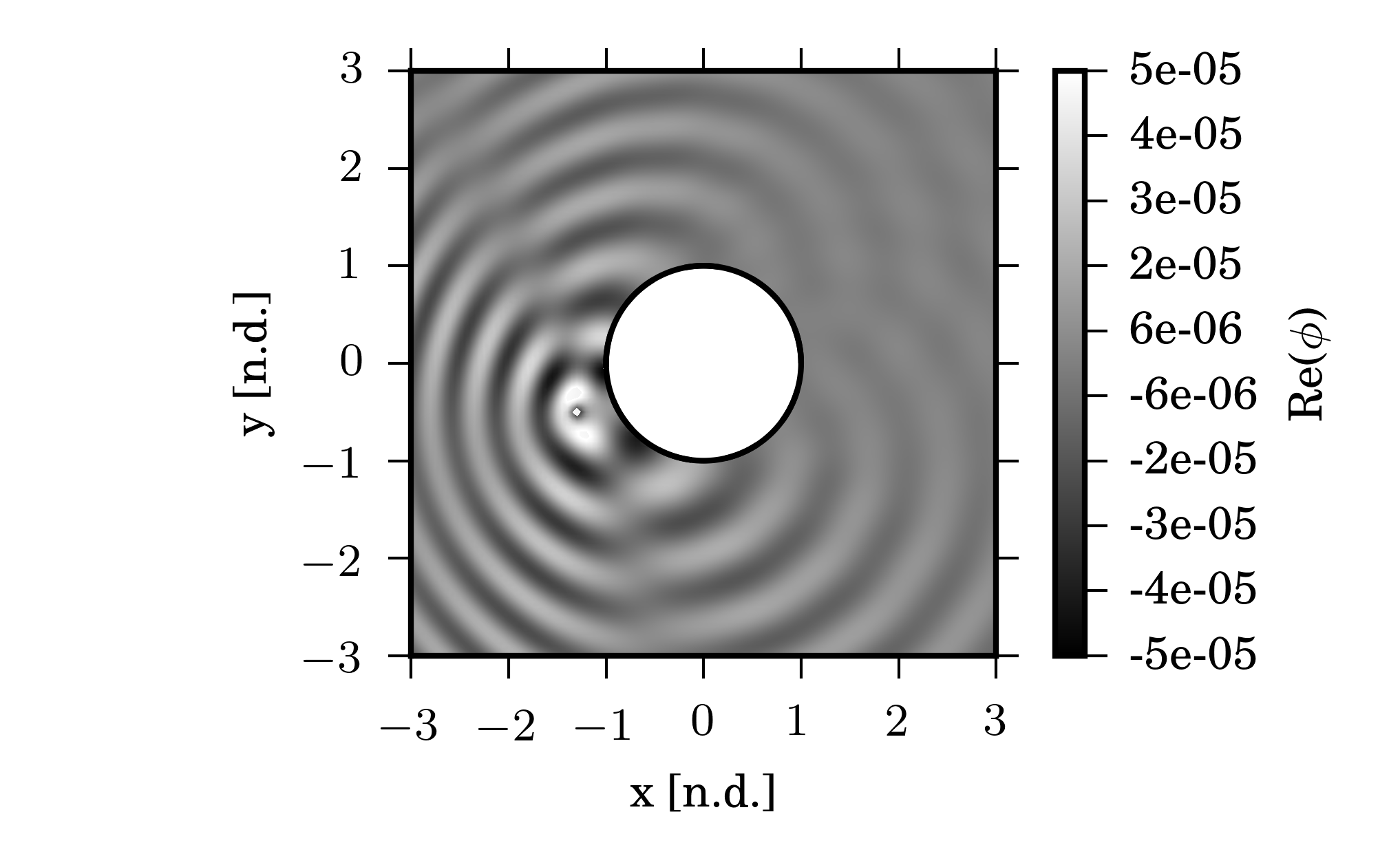}
	}
	\hfill
	\subfloat[\label{fig:Figure_14b}]{%
		\includegraphics[trim=1.3cm 0cm 0.3cm 0cm, clip=true,width=0.32\textwidth]{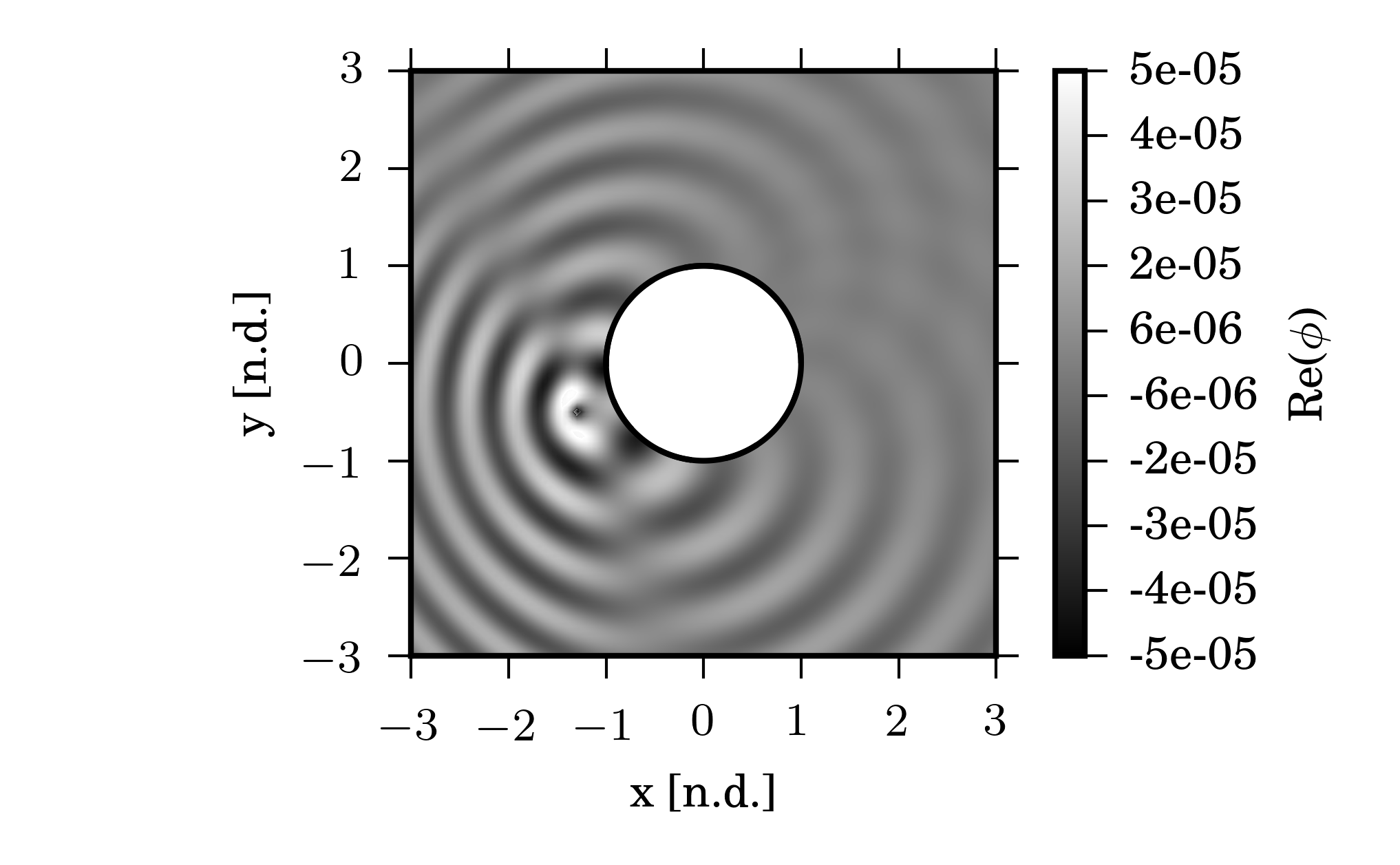}
	}
	\hfill
	\subfloat[\label{fig:Figure_14c}]{%
		\includegraphics[trim=1.3cm 0cm 0.3cm 0cm, clip=true,width=0.32\textwidth]{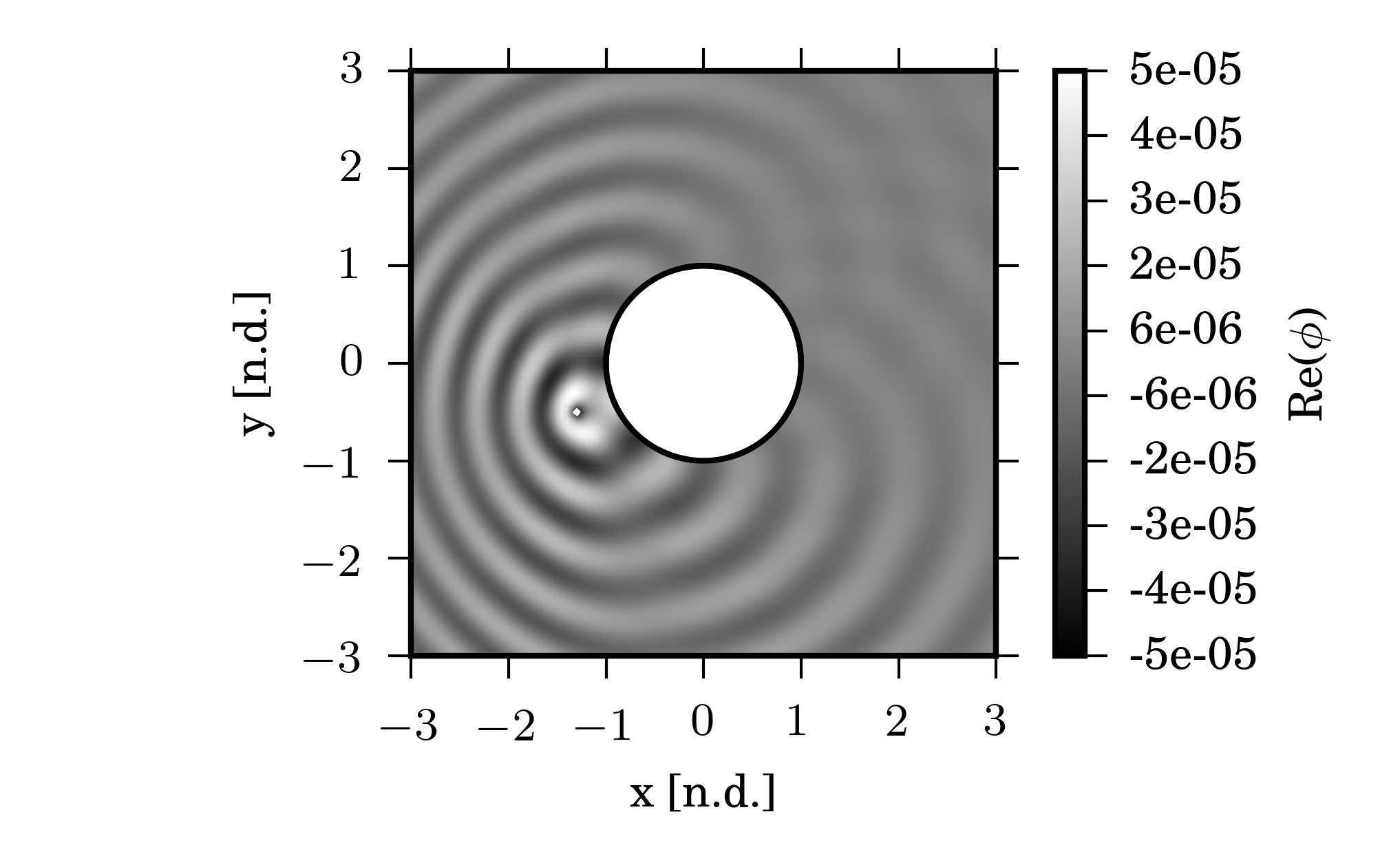}
	}
	\caption{\footnotesize  Real part of the acoustic velocity potential, $Re(\phi)$, at a non-dimensional frequency $ka=10$ for $M_\infty=0.3$. The BE solution of the weakly non-uniform potential flow Helmholtz equation (a), the FE solution of the full potential linearized Helmholtz equation (b) and the BE solution of the uniform flow Helmholtz equation~\cite{Wu1994} (c) are shown for the problem in Fig.~\ref{fig:geometry_BEM} but locating a monopole point source at ${\bf{x}}_s=(-1.3a,-0.5a)$.}
	\label{fig:dummy_test2}
\end{figure}

The previous results have been presented in terms of computed value of the acoustic velocity potential. The acoustic pressure $p$ is another quantity of practical interest. The absolute value of the acoustic pressure at $r_{fp}=4a$ for $M_\infty=0.1$ and $0.3$ is shown in Figs.~\ref{fig:BEM_Integ_Taylor_Lorentz_accuracy_dBPressM01} and \ref{fig:BEM_Integ_Taylor_Lorentz_accuracy_dBPressM03}. The reference solution is compared to the approximate integral solutions at $ka=10$ for different observer angular positions (the angle $\theta$ measured from the $x$-axis and positive counterclockwise). For $M_\infty=0.1$, all of the integral formulations are mostly within $1$~dB and $5$~dB of the reference solution. However, at $M_\infty=0.3$ the error incurred in the Taylor-Helmholtz solution reaches $10$ dB. Nonetheless, the BE prediction based on wave propagation on uniform flows overestimates the reference solution in the shielded area [$\theta=30^\circ-70^\circ$], where the incident and the scattered field interfere destructively and the flow is not aligned with the uniform stream. Note that the uniform flow Helmholtz formulation neglects wave refraction effects due to non-uniformities.

\begin{figure}[t]
	\centering
	\includegraphics{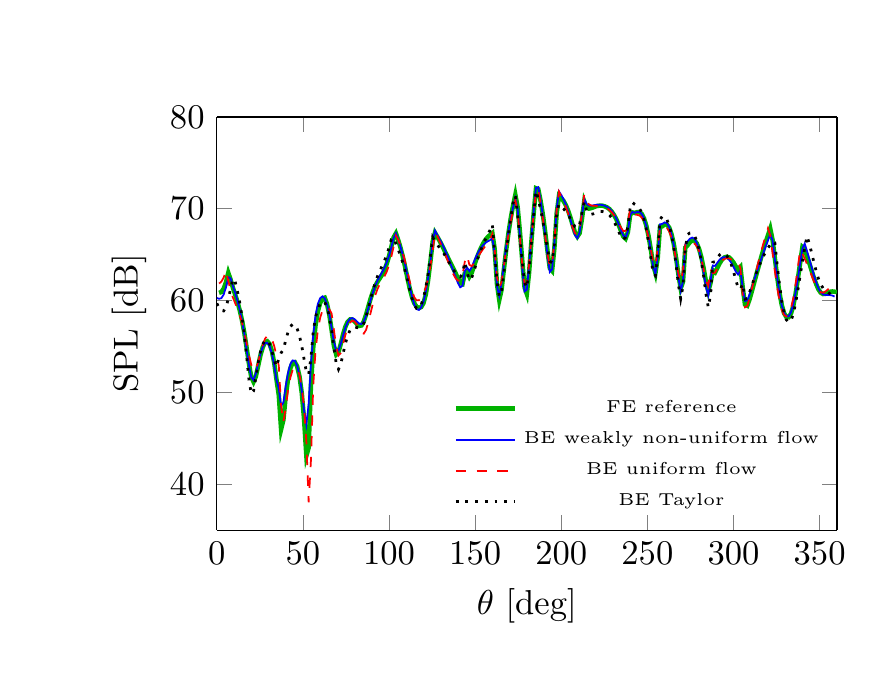} 
	\caption{\footnotesize Acoustic pressure directivity along a field point circular arc with radius $r_{fp}=4a$ at a non-dimensional frequency $ka=10$ for a mean flow Mach number $M_\infty=0.1$. The BE solutions (see Fig.~\ref{fig:geometry_BEM}) are based on the integral formulations Eq.~(\ref{eq:Non_uniform_boundary_integral_surface_v3}) (BE weakly non-uniform), Wu and Lee~\cite{Wu1994} (BE uniform flow) and Eq.~(\ref{eq:Taylor_boundary_integral_surface}) (BE Taylor).}
	\label{fig:BEM_Integ_Taylor_Lorentz_accuracy_dBPressM01}
\end{figure}
\begin{figure}[t]
	\centering
	\includegraphics{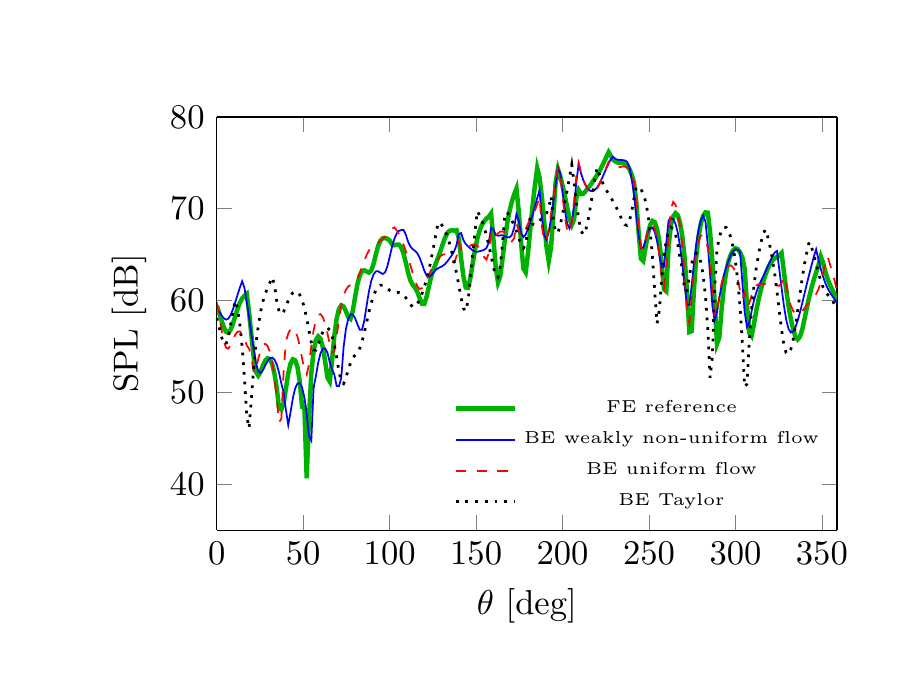} 
	\caption{\footnotesize Acoustic pressure directivity along a field point circular arc with radius $r_{fp}=4a$ at a non-dimensional frequency $ka=10$ for a mean flow Mach number $M_\infty=0.3$. The BE solutions (see Fig.~\ref{fig:geometry_BEM}) are based on the integral formulations Eq.~(\ref{eq:Non_uniform_boundary_integral_surface_v3}) (BE weakly non-uniform), Wu and Lee~\cite{Wu1994} (BE uniform flow) and Eq.~(\ref{eq:Taylor_boundary_integral_surface}) (BE Taylor).}
	\label{fig:BEM_Integ_Taylor_Lorentz_accuracy_dBPressM03}
\end{figure}

For the weakly non-uniform potential flow Helmholtz equation the error varies linearly with frequency (see Fig.~\ref{fig:BEM_Non_Uni_indErr_freqSens}) and with $M^2_\infty$ (see Fig.~\ref{fig:BEM_Integ_Taylor_Lorentz_accuracy_L2err}). The error is independent of the observer distance when it is sampled on a region where the flow is almost uniform (see Fig.~\ref{fig:BEM_Non_Uni_indErr_Unif_flow}) because the formulation is exact for a uniform flow, namely no additional error is generated as the field point radius increases. These results validate further the error analysis of Section~\ref{sec:Error_Estimate_small_perturbation_convected_wave_equation}.

As in the previous test case, the weakly non-uniform flow Helmholtz formulation outperforms both the uniform flow Helmholtz and the Taylor-Helmholtz solutions (see Figs.~\ref{fig:BEM_Integ_Taylor_Lorentz_accuracy_L2err}, \ref{fig:BEM_Integ_Taylor_Lorentz_accuracy_dBPressM01} and \ref{fig:BEM_Integ_Taylor_Lorentz_accuracy_dBPressM03}). The improvement against the uniform flow Helmholtz solution is larger at low Mach numbers. Since $M'_0\sim M_\infty$, for increasing $M_\infty$ the error term of order $M_\infty M'_0/L_A^2$ in Eq.~(\ref{eq:erorr_weakly_perturbed_potent_Helmholtz}), is significant both for the uniform and weakly non-uniform flow solutions, eventually reducing the improvement given by the non-uniform flow formulation on the uniform flow approximation.

\section{Concluding remarks}

A novel integral formulation for sound radiation on non-uniform mean flows has been proposed. It provides an approximate solution of the full potential linearized Helmholtz equation by assuming a weakly non-uniform potential mean flow. A Green's function, describing non-uniform flow effects on wave propagation, was derived for potential subsonic mean flows. The key advantage is that the Green's function and subsequent integral equation are presented in the physical space. The integral formulation is exact for wave propagation on a uniform flow and the error generated in the non-uniform flow region remains constant when wave propagation occurs on a uniform flow. This distinguishes the current approach from the \emph{Taylor transformation} integral formulation applied by a number of previous researchers in computational Aeroacoustics~\cite{Astley1986,Agarwal2007,Mayoral2013}.

\par A comparison with the Taylor integral formulation and the integral solution for the uniform flow Helmholtz equation (Lorentz formulation) has been presented. First, an analysis based on wave extrapolation on a non-uniform flow was performed. In this case, the condition that $M_0'\ll M_\infty$ was satisfied at all points outside the integral surface. For the weakly non-uniform potential flow Helmholtz solution the error varies linearly with $M_\infty$ and is proportional to the frequency $f$. The new approach performs better than the Taylor formulation in which the error varies as $M_\infty^2$ and is linear in $f$. It also outperforms the uniform flow Helmholtz equation, reducing the error of $40\%$ for $M_\infty \leq 0.3$.

\par In a second test case the acoustic field was solved up to the solid scattering surface by means of a full BE approach. On the boundary surface itself $M_0'$ is of the same order of magnitude of $M_\infty$, pushing to the limit the underlying assumption of the formulation of the analysis $M_0'\ll M_\infty$. An error analysis on the integral solution based on the weakly non-uniform potential flow Helmholtz equation shows that the $L^2$ error is proportional to $M^2_\infty$ and varies linearly with $f$. A consistent improvement in accuracy is achieved against the integral solutions based on uniform mean flows and on the Taylor-Helmholtz equation for a non-uniform flow. A significant advantage of the current weakly non-uniform potential flow Helmholtz solution over the uniform flow Helmholtz solution is evident at low Mach numbers, where for instance a reduction of $50\%$ of the error  is observed at $M_\infty=0.1$.

\section*{Acknowledgments}
The authors gratefully acknowledge the European Commission for the support of the project FP7-PEOPLE-2013-ITN CRANE, Grant Agreement 606844. The third author wishes to thank the Royal Commission for the Exhibition of 1851 for their outstanding support. Alastair Gregory kindly provided his results on the Taylor transformation to validate the Taylor-Helmholtz formulation.

 \appendix

\section{Integral formulation along the boundary surface}\label{sec:Integral_form_along_bound_surface}

An extension of Eq.~(\ref{eq:Non_uniform_boundary_integral_domain}) to the boundary surface is given on the basis of a limit approach~\cite{Wu2000}, following Wu and Lee~\cite{Wu1994}. When an exterior wave propagation problem is considered, the domain $\Omega$ (see Fig.~\ref{fig:geometry_Domain}) is modified by subtracting a hemisphere of radius $\epsilon$ in the neighbourhood of ${\bf{x}}_p \in \Omega$,
\begin{equation}
\begin{split}
\phi({\bf{x}}_p) &= \int_{\partial \Omega+\partial \Omega_\epsilon}\left( G \frac{\partial \phi}{\partial n}- \phi\frac{\partial G}{\partial n}\right) dS\\
&-\int_{\partial \Omega+\partial \Omega_\epsilon} \left[2\mathrm{i}k{\bf{M}}_0\cdot {\bf{n}} G \phi + M^2_\infty\left(G \frac{\partial \phi}{\partial x}-\phi \frac{\partial G}{\partial x}\right)n_x\right] dS.
\end{split}
\label{eq:Non_Uniform_Boundary_Integral_Equation}
\end{equation}
When the radius $\epsilon$ tends to zero, the surface of the hemisphere tends to zero as $\epsilon^2$. Since $G$ varies as $1/\epsilon$, the contribution of the terms including $G$ to the integral over the infinitesimal surface is zero. Hence, from Eq.~(\ref{eq:Non_Uniform_Boundary_Integral_Equation}) one has
\begin{equation}
\begin{split}
\phi({\bf{x}}_p) &= \int_{\partial \Omega} \left(G \frac{\partial \phi}{\partial n}- \phi\frac{\partial G}{\partial n}\right)dS\\
&-\int_{\partial \Omega} \left[2\mathrm{i}k{\bf{M}}_0\cdot {\bf{n}} G \phi + M^2_\infty\left(G \frac{\partial \phi}{\partial x}-\phi \frac{\partial G}{\partial x}\right)n_x\right] dS\\
& - \phi({\bf{x}}_p)\int_{\partial \Omega_\epsilon}\left( \frac{\partial G}{\partial n}-M_\infty^2\frac{\partial G}{\partial x}n_x\right) dS.
\end{split}
\label{eq:Non_uniform_boundary_integral_surface_v1}
\end{equation}
In the limit of $\epsilon \rightarrow 0$ the Green's function in Eq.~(\ref{eq:convected_small_distur_Helmholtz_adjoint_operatror_eq}) tends to the Green's function of the static operator~\cite{Wu1994} $G_0$. The static operator associated with Eq.~(\ref{eq:convected_small_distur_Helmholtz_adjoint_operatror_eq}) is given by  
\begin{equation}
\nabla^2G_0 - M^2_\infty \frac{\partial^2 G_0}{\partial x^2} = -\delta({\bf{x}}_p-{\bf{x}}).
\label{eq:static_operator_smallPertConvHelm}
\end{equation}	
The solution of the equation above for a free field problem gives $G_0({\bf{x}}_p,{\bf{x}}) = 1/(4\pi R_M)$. Wu and Lee~\cite{Wu1994} have shown that the integral on the hemisphere of radius $\epsilon$ can be rewritten on the boundary surface $\partial \Omega$ as follows:
\begin{equation}
\int_{\partial \Omega_\epsilon}\left( \frac{\partial G_0}{\partial n}-M_\infty^2\frac{\partial G_0}{\partial x}n_x\right)dS = -\int_{\partial \Omega} \left(\frac{\partial G_0}{\partial n}-M_\infty^2\frac{\partial G_0}{\partial x}n_x\right)dS.
\label{eq:relation_surface_cauchy_value}
\end{equation}
Therefore, applying Eq.~(\ref{eq:relation_surface_cauchy_value}) to Eq.~(\ref{eq:Non_uniform_boundary_integral_surface_v1}) yields	
\begin{equation}
\begin{split}
\hat C({\bf{x}}_p)\phi({\bf{x}}_p) &= \int_{\partial \Omega} \left(G \frac{\partial \phi}{\partial n}- \phi\frac{\partial G}{\partial n}\right) dS \\
&-\int_{\partial \Omega} \left[ 2\mathrm{i}k{\bf{M}}_0\cdot {\bf{n}} G \phi + M^2_\infty\left(G \frac{\partial \phi}{\partial x}-\phi \frac{\partial G}{\partial x}\right)n_x\right] dS ,
\end{split}
\end{equation}
where
\begin{equation}
\hat C({\bf{x}}_p)=
\left\{
\begin{array}{lr}
1 & {\bf{x}}_p\in\Omega \\
1 -\int_{\partial \Omega}\left(\frac{\partial G_0}{\partial n}-M_\infty^2\frac{\partial G_0}{\partial x}n_x \right)dS & \quad {\bf{x}}_p\in\partial\Omega\\
\end{array}\right.
\end{equation}

\medskip

\section{2D Green's function}

This section presents the 2D Green's function for wave propagation on a weakly non-uniform potential mean flow. The formulation is limited to subsonic flows. The solution is based on the proof given in Section~\ref{sec:Green_function_adjoint_operator}. The Green's function for the Taylor-Helmholtz formulation is also derived.

The Green's function for the Helmholtz operator in the Taylor-Lorentz space, Eq.~(\ref{eq:Helmholtz_Equation_Lorentz_Taylor}), is given by
\begin{equation}
G(\tilde {\bf{X}})=\frac{\mathrm{i}}{4 \beta_\infty} H_0^{(2)}\left(\tilde k \tilde R_{2D}\right),
\end{equation}
where $\tilde R_{2D}=\sqrt{\tilde X^2+ \tilde Y^2 }$, $H_0^{(2)}$ is the Hankel function of the second type of order zero and  $\tilde k = k/\beta_\infty$. A time harmonic solution is sought for $G$, namely $G(\tilde {\bf{X}},\tilde T)=G(\tilde {\bf{X}})e^{i \tilde \omega \tilde T}$. The inverse Taylor-Lorentz transformation is applied to the above equation retrieving the Green's function in the physical space, as described in Section~\ref{sec:Green_function_adjoint_operator},
\begin{equation}
G({\bf{x}})=\frac{\mathrm{i}}{4 \beta_\infty} H_0^{(2)}\left(\frac{k\sqrt{x^2+\beta_\infty^2y^2}}{\beta_\infty^2}\right)\mathrm{e}^{-\mathrm{i} k\left(\frac{M_\infty x}{\beta^2_\infty}+\frac{\Phi_0'({\bf{x}})}{c_\infty} \right)}.
\end{equation}
Extending the Green's function to a generic source position ${\bf {x}}_s=(x_s,y_s)$ yields
\begin{equation}
G({\bf{x}},{\bf{x}}_s)=\frac{\mathrm{i}}{4 \beta_\infty} H_0^{(2)}\left(\frac{kR_{M,2D}}{\beta_\infty^2}\right)\mathrm{e}^{-\mathrm{i} k\left(\frac{M_\infty (x-x_s)}{\beta^2_\infty}+\frac{\Phi_0'({\bf{x}})-\Phi_0'({\bf{x}}_s)}{c_\infty} \right)}
\label{eq:Green_function_physical_space_Taylor_Lorentz_adjoint2D}
\end{equation}
where $R_{M,2D}=\sqrt{(x-x_s)^2 +\beta_\infty^2(y-y_s)^2}$. The Green's function associated with the static operator Eq.~(\ref{eq:static_operator_smallPertConvHelm}) is $G_0=-log(R_{M,2D})/2\pi$. 

For a uniform flow, Eq.~(\ref{eq:Green_function_physical_space_Taylor_Lorentz_adjoint2D}) reduces to the expression provided by Bailly and Juv\'e~\cite{Bailly2000} including a reverse mean flow. On the other hand, if $M^2_\infty\ll 1$, the 2D Green's function can be rewritten for the Taylor-Helmholtz formulation as
\begin{equation}
G_{T}({\bf{x}},{\bf{x}}_s)=\frac{\mathrm{i}}{4 \beta_\infty} H_0^{(2)}(kR_{2D})\mathrm{e}^{-\mathrm{i} k\frac{\Phi_0({\bf{x}})-\Phi_0({\bf{x}}_s)}{c_\infty}}
\end{equation}
where $R_{2D} = \sqrt{(x-x_s)^2+(y-y_s)^2} $.





\end{document}